\documentclass[numberedappendix]{emulateapj}
\usepackage{amsmath}

\slugcomment{ApJ, accepted; \today}

\newcommand{\G}{\Gamma}

\newcommand{\sT}{\sigma_{\rm T}}
\newcommand{\p}{^\prime}
\newcommand{\e}{\epsilon}
\newcommand{\g}{\gamma}
\newcommand{\gp}{\gamma^{\prime}}

\newcommand{\tp}{t^\prime}
\newcommand{\ep}{\epsilon^\prime}

\newcommand{\dD}{\delta_{\rm D}}

\newcommand{\psim}{\lower.5ex\hbox{$\; \buildrel \propto \over\sim \;$}}
\newcommand{\lbar}{\lower.0ex\hbox{$\; \buildrel
{\lower0.0ex \hbox{-}} \over\lambda  \;$}}

\newcommand{\dotg}{\dot{\g}}
\newcommand{\tilN}{\tilde{N}}
\newcommand{\tilQ}{\tilde{Q}}
\newcommand{\tilF}{\tilde{F}}
\newcommand{\term}{ \left( \frac{1}{t_{\rm esc}}-i\omega \right) }
\newcommand{\termb}{ \left( \frac{1}{t_{\rm esc}^2}+\omega^2 \right) }
\newcommand{\fpr}{f^\prime}

\newcommand{\cm}{\mathrm{cm}}

\newcommand{\erg}{\mathrm{erg}}

\newcommand{\MeV}{\mathrm{MeV}}
\newcommand{\GeV}{\mathrm{GeV}}
\newcommand{\TeV}{\mathrm{TeV}}
\newcommand{\s}{\mathrm{s}}

\newcommand{\ddays}{\mathrm{days}}

\newcommand{\Hz}{\mathrm{Hz}}

\newcommand{\Gauss}{\mathrm{G}}

\newcommand{\fermi}{{\em Fermi}}

\shorttitle{Fourier Analysis of Blazar Variability}
\shortauthors{Finke \& Becker}

\begin{document}
\title{Fourier Analysis of Blazar Variability}

\author{Justin D.\ Finke$^1$ and Peter A.\ Becker$^2$}


\affil{$^1$U.S.\ Naval Research Laboratory, Code 7653, 4555 Overlook Ave.\ SW,
        Washington, DC,
        20375-5352\\
	$^2$School of Physics, Astronomy, and Computational Sciences, MS 5C3, 
	George Mason University, 4400 University Drive, Fairfax, VA 
}

\email{justin.finke@nrl.navy.mil}

\begin{abstract}

Blazars display strong variability on multiple timescales and in
multiple radiation bands. Their variability is often characterized by
power spectral densities (PSDs) and time lags plotted as functions of
the Fourier frequency.  We develop a new theoretical model 
based on the analysis of the electron transport (continuity)
equation, carried out in the Fourier domain. The continuity equation
includes electron cooling and escape, and a derivation of the emission
properties includes light travel time effects associated with a
radiating blob in a relativistic jet. The model successfully
reproduces the general shapes of the observed PSDs and predicts
specific PSD and time lag behaviors associated with variability in the
synchrotron, synchrotron self-Compton (SSC), and external Compton (EC)
emission components, from sub-mm to $\gamma$-rays.  We discuss
applications to BL Lacertae objects and to flat-spectrum radio quasars
(FSRQs), where there are hints that some of the predicted
features have already been observed. We also find that FSRQs should
have steeper PSD power-law indices than BL Lac objects at Fourier
frequencies $\la 10^{-4}$\ Hz, in qualitative agreement with
previously reported observations by the {\em Fermi} Large Area
Telescope.

\end{abstract}

\keywords{BL Lacertae objects:  general --- quasars:  general --- 
radiation mechanisms:  nonthermal --- galaxies: active --- 
galaxies:  jets}

\section{Introduction}

Blazars, active galactic nuclei (AGN) with jets moving at relativistic
speeds aligned with our line of sight, are the most plentiful of the
identified point sources observed by the {\em Fermi} Large Area
Telescope \citep[LAT;][]{abdo10_1fgl,nolan12_2fgl}.  Their spectral
energy distributions (SEDs) are dominated by two components.  At lower
frequencies there is a component produced via synchrotron
emission, peaking in the radio through X-rays.  There is also a high
energy component in $\g$-rays, most likely due to
Compton-scattered emission.  The seed photons for Compton scattering
could be the synchrotron photons themselves \citep[synchrotron
self-Compton or SSC;][]{bloom96}, or they could be produced outside
the jet (external Compton or EC) by the accretion disk
\citep{dermer93,dermer02}, the broad line region \citep{sikora94}, or
a dust torus \citep{kataoka99,blazejowski00}.  The Doppler-boosting
due to the combination of relativistic speed and a small
jet inclination angle amplifies the observed flux,  shifting the
emission to higher frequencies, and decreasing the variability
timescale.

The LAT monitors the entire sky in high-energy $\g$-rays every 3
hours, providing well-sampled light curves of blazars on long time
scales. For some sources, $\g$-ray data have been supplemented by high
cadence observations in the radio through very-high energy (VHE)
$\gamma$-rays creating unprecedented light curves with few gaps in
wavelength or time
\citep[e.g.,][]{abdo11_mrk421,abdo11_mrk501}. Blazar variability is
often characterized by power spectral densities \citep[PSDs;
e.g.,][]{abdo10_var,chatterjee12,hayashida12,nakagawa13,sobolewska14},
which are essentially representations of the Fourier transform
without phase information. Although the LAT can provide long baseline,
high cadence light curves, it has difficulty probing short timescales
for all but the brightest flares. However, shorter-timescale
variability may be observed with optical, X-ray or VHE instruments
\citep[e.g.,][]{zhang99,kataoka01,zhang02_2155,cui04,aharonian07_2155,rani10}.
Less often, Fourier-frequency dependent time lags between two energy
channels are computed from light curves
\citep[e.g.,][]{zhang02_mrk421}.  The PSDs of light curves at
essentially all wavelengths resemble power-laws in frequency,
$S(f)\propto f^{-b}$, with typically $b\sim 1-3$, usually steeper than
PSDs found from Seyfert galaxies \citep{kataoka01}. Despite the
popularity of PSDs for characterizing variability, their theoretical
motivation has not been thoroughly explored \citep[although
see][]{mastichiadas13}.

In this paper, our goal is to bridge the gap between theory and
observations by exploiting a powerful new mathematical approach for
the modeling and interpretation observed PSDs and time lags.
Theoretically, the variability of blazars is often described by a
continuity equation
\citep[e.g.,][]{chiaberge99,li00,boett02a,chen11,chen12}.  This
equation describes the evolution of electrons in a compact region of
the jet, which is homogeneous by assumption.  The electrons are
injected as a function of time and energy and the electron
distribution evolves due to energy loss and escape.  The expected
electromagnetic emission observers might detect can be compared with
observations \citep[e.g.,][]{boett04,joshi07}.  This can allow one to
explore individual flares; however, in this paper, we take a different
approach, by studying the electron continuity equation in the
{\em Fourier domain}.  This allows the exploration of individual
flares, as well as the study of aggregated long-timescale variability
of sources using Fourier transform-related quantities such as PSDs and
phase and time lags.  We focus here on long-timescale
variability, including multiple epochs of flaring and quiescence.  A
similar study applying the same Fourier transform concept to the
modeling of time lags in accreting Galactic black hole candidates has
been recently carried out by \citet{kroon14}.

We begin in Section \ref{definition}, by defining the Fourier
transform and its inverse, and various other functions
used throughout the rest of the paper.  In Section
\ref{continuity_eqn}, we explore analytic solutions to the continuity
equation in the Fourier domain, and present solutions in terms of
PSDs.  In Section \ref{lagsection}, we explore the solutions in terms
of time lags between different electron Lorentz factor ``channels'' as
a function of Fourier frequency.  We explore the expected synchrotron,
SSC, and EC PSDs and Fourier frequency-dependent time lags in Section
\ref{emission}, including light travel time effects due to the
emitting region's finite size.  We discuss applications to some PSDs
and time lags in the literature in Section \ref{applications}, and
conclude with a discussion of our simplifying assumptions and
observational prospects in Section \ref{discussion}.  Several of the
detailed derivations are relegated to the appendices.

\section{Definitions}
\label{definition}

Several definitions of the Fourier transform and associated quantities
are used throughout the literature.  Here we make the definitions used
in this paper explicit.  For a real function $x(t)$, we define the
Fourier transform by
\begin{eqnarray}
\label{FT_define}
\tilde{x}(f) = \int^{\infty}_{-\infty} dt\ x(t)\ e^{2\pi i f t}\ = 
\int^{\infty}_{-\infty} dt\ x(t)\ e^{i \omega t}\ ,
\end{eqnarray}
where $i^2=-1$. We will indicate the Fourier transform by a tilde, the
Fourier frequency by $f$, and the angular frequency by $\omega=2\pi
f$.  We define the inverse Fourier transform by
\begin{eqnarray}
\label{IFT_define}
x(t) = \int^{\infty}_{-\infty} df\ \tilde{x}(f)\ e^{ -2\pi i f t } = 
\frac{1}{2\pi}\int^{\infty}_{-\infty}d\omega\ \tilde{x}(\omega)\ e^{-i \omega t}\ .
\end{eqnarray}
We define the PSD
\begin{eqnarray}
S(f) = |\tilde{x}(f)|^2 = \tilde{x}(f)\tilde{x}^*(f)\ 
\end{eqnarray}
where the asterisk indicates the complex conjugate.
We will make use of the related
representation of the Dirac $\delta$-function
\begin{eqnarray}
\label{dirac}
\delta(t-t_0) = \int^{\infty}_{-\infty} df\ e^{ 2\pi i f (t-t_0)}\ .
\end{eqnarray}
We will use two versions of the Heaviside function, a 
``step'' function defined by
\begin{equation}
H(x) = \left\{ \begin{array}{ll}
1 & x > 0 \\
0 & \mathrm{otherwise}
\end{array}
\right. \ ,
\end{equation}
and the two-sided Heaviside ``top-hat'' function, 
\begin{equation}
H(x; a, b) = \left\{ \begin{array}{ll}
1 & a < x <b \\
 0 & \mathrm{otherwise}
\end{array}
\right. \ .
\end{equation}
In Section \ref{generalsoln} we will use the lower 
incomplete Gamma function given by
\begin{eqnarray}
\g_g(a,x) = \int_0^x dy\ y^{a-1}e^{-y}\ .
\end{eqnarray}

\section{Continuity Equation in Fourier Space}
\label{continuity_eqn}

\subsection{General Solution}

The evolution of electrons in a nonthermal plasma ``blob'' can be
described by a continuity equation given by
\citep[e.g.,][]{chiaberge99,li00,boett02a,chen11,chen12}
\begin{eqnarray}
\frac{\partial N_e}{\partial t} + \frac{\partial}{\partial\g}[ \dotg(\g,t) N_e(\g; t) ] +
\frac{N_e(\g; t)}{t_{\rm esc}(\g,t)} = Q(\g, t) \ ,
\end{eqnarray}
where $N_e(\g;t)\ d\gamma$ gives the number of electrons with Lorentz
factor between $\gamma$ and $\gamma+d\gamma$ at time $t$.  Here
$\dotg(\g,t)$ is the rate at which electrons lose or gain energy,
$t_{\rm esc}(\g,t)$ is the escape timescale, and $Q(\g,t)$ is the rate
at which electrons are injected in the jet.  In the simple model
presented here, we will assume that size of the plasma blob does not
change with time, so that adiabatic losses can be neglected.  We
will also assume that the electron distribution in the blob is
homogeneous and isotropic, and that variations occur throughout the
blob simultaneously. This is a common and useful assumption, although
it is somewhat unphysical, as in order for the blob to be
causally connected, variations cannot propagate through the blob
faster than the speed of light $c$.

Assuming that $\dotg$ and $t_{\rm esc}$ are independent of $t$,
we can take the Fourier transform of both sides of the
continuity equation leading to
\begin{eqnarray}
\label{fourier_cont_eqn}
-2\pi if \tilN_e(\g,f) + \frac{\partial}{\partial\g} [\dotg(\g) \tilN_e(\g,f)] + 
\frac{\tilN_e(\g,f)}{t_{\rm esc}(\g)} 
\nonumber \\
= \tilQ(\g, f)\ 
\end{eqnarray}
where $\tilQ(\g, f)$ is the Fourier transformed source term.  This is
a linear ordinary first-order differential equation with a relatively
simple solution.  It is shown in Appendix \ref{FTCEsoln} that if
$\dotg\le 0$
\begin{eqnarray}
\label{Nsoln1}
\tilN_e(\g,f) = \frac{1}{|\dotg(\g)|}\int_\g^{\infty} d\gp\ \tilQ(\gp,f)\ 
\nonumber \\ \times
\exp\left[ -\int_{\g}^{\gp} \frac{d\g^{\prime\prime}}{|\dotg(\g^{\prime\prime})|}
\left(\frac{1}{t_{\rm esc}(\g^{\prime\prime})} - i\omega\right)\right]\ .
\end{eqnarray}
If $t_{\rm esc}$ is independent of $\g$ and cooling is from
synchro-Compton processes, so that $\dotg = -\nu\g^2$, then one can perform the integral in the 
exponent, and
\begin{eqnarray}
\label{Nsoln2}
\g^2 \tilN_e(\g,f) = \frac{1}{\nu}\ \exp\left[ \frac{-1}{\nu\g}
\left( \frac{1}{t_{\rm esc}}-i\omega \right) \right]
\nonumber \\  \times
\int_{\g}^{\infty} d\gp\ \tilQ(\gp,f)\  
\exp\left[ \frac{1}{\nu \gp} \left( \frac{1}{t_{\rm esc}}-i\omega \right) \right]\ .
\end{eqnarray}

\subsection{Green's Function Solution}

Consider an instantaneous injection of monoenergetic
electrons with Lorentz factor $\g_0$ at $t=0$.  Then $\tilQ(\g,f) =
Q_0\delta(\g-\g_0)$ in Equation (\ref{Nsoln2}) and one gets
\begin{flalign}
\g^2 \tilN_e(\g,f) & = \frac{Q_0}{\nu}\ 
\exp\left[ \frac{-1}{\nu}
\left( \frac{1}{\g} - \frac{1}{\g_0}\right)
\left( \frac{1}{t_{\rm esc}}-i\omega \right) \right] 
 \nonumber \\ & \times
H(\g_0-\g)\ .
\end{flalign}
The PSD is
\begin{flalign}
S(\g,f) & = |\g^2 \tilN_e(\g,f)|^2 
\nonumber \\ & 
= \left[\frac{Q_0}{\nu}\right]^2
\exp\left[ \frac{-2}{\nu t_{\rm esc}}
\left( \frac{1}{\g} - \frac{1}{\g_0}\right)\right]
H(\g_0-\g)\ .
\end{flalign}
The result is white noise for all electron Lorentz factors ($\g$) and
Fourier frequencies ($f$).  

\subsection{Colored Noise}

Since the PSDs of blazars resemble colored noise, and electrons are
generally thought to be injected as power laws in $\g$, one might
expect that
\begin{flalign}
\label{tilQ}
\tilQ(\g,f) & = Q_0 (f/f_0)^{-a/2}\g^{-q}
\nonumber \\ & \times
H(\g;\g_1,\g_2) H(f; f_1, f_2)
\end{flalign}
where $f_0$ is some constant frequency and $a\ge0$.  That is, in the
jet, shocks will occur randomly which accelerate and inject particles
as a power-law distribution in $\g$ between $\g_1$ and
$\g_2$ with index $q$. We will deal only with frequencies
in the range $f_1 \le f \le f_2$.  These limits are needed for the PSD
to be normalized to a finite value.  Frequencies greater than the
inverse of the blob's light crossing timescale are particularly
unphysical, although we allow this for two reasons.  First, it allows
us to compare with other theoretical studies that allow variations
faster than the light crossing timescale
\citep[e.g.,][]{chiaberge99,zacharias13}.  Second, our blob is already
unphysical, since we allow variations throughout the blob
simultaneously in the blob's comoving frame.  The normalization
constant is related to the time-averaged power injected in electrons
$\langle L_{\rm inj}\rangle$ over a time interval $\Delta t$ by
\begin{equation}
\label{Q0}
Q_0 = \frac{2\pi \Delta t \langle L_{\rm inj}\rangle}{m_e c^2 G
\sqrt{ I_r^2 + I_i^2 - 2 I_r I_0 + I_0^2} }\ .
\end{equation}
A derivation of this equation and definitions of the quantities $G$,
$I_r$, $I_i$, and $I_0$ can be found in Appendix
\ref{normalization}. With $\tilQ(\g,f)$, given by Equation
(\ref{tilQ}), Equation (\ref{Nsoln2}) can be rewritten as
\begin{flalign}
\label{Nsoln3}
\g^2\tilN_e(\g,f) & = 
Q_0 (f/f_0)^{-a/2} \exp\left[ \frac{-1}
{\nu\g}\term \right] \nu^{q-2}
\nonumber \\ & \times 
\term^{1-q}
\int_{u_{\min}}^{u_{\max}} du\ u^{q-2}\ e^{u}\ 
\end{flalign}
where 
\begin{eqnarray}
u_{\min} = \frac{1}{\nu\g_2}\term\ 
\end{eqnarray}
and
\begin{eqnarray}
u_{\max} = \frac{1}{\nu\max(\g,\g_1)}\term\ .
\end{eqnarray}

\subsection{Electron Injection Index  $q=2$ }

It is instructive to look at the case where $q=2$.  In this case, the
remaining integral in Equation (\ref{Nsoln3}) can easily be performed
analytically.  Then
\begin{flalign}
\label{Nsoln4}
\g^2\tilN_e(\g,f) & = \frac{Q_0 (f/f_0)^{-a/2}}{1/t_{\rm esc}-i\omega}
\exp\left[ \frac{-1}{\nu\g}\left( \frac{1}{t_{\rm esc}}-i\omega\right)\right]
\nonumber \\ & \times 
\left[ e^{u_{\max}}-e^{u_{\min}} \right]\ ,
\end{flalign}
and the PSD is
\begin{flalign}
\label{PSD_eqn2}
S(\g,f) & = |\g^2\tilN_e(\g,f)|^2  
\nonumber \\ &
= \exp\left[\frac{-2}{\nu\g t_{\rm esc}}\right] 
Q_0^2 (f/f_0)^{-a}\termb^{-1}
\nonumber \\ & \times 
\Biggr\{ \exp\left[ \frac{2}{\nu\g_2 t_{\rm esc}} \right] + 
\exp\left[ \frac{2}{\nu\max(\g,\g_1) t_{\rm esc}} \right]  
\nonumber \\ & 
-2\exp\left[ \frac{1}{\nu t_{\rm esc}}\left( \frac{1}{\max(\g,\g_1)} + \frac{1}{\g_2}\right)  \right]
\nonumber \\ & \times
\cos\left[  \frac{\omega}{\nu}    \left( \frac{1}{\max(\g,\g_1)} - \frac{1}{\g_2} \right)  \right]
\Biggr\}\ .
\end{flalign}

\begin{figure}
\vspace{5.2mm} 
\epsscale{1.0} 
\plotone{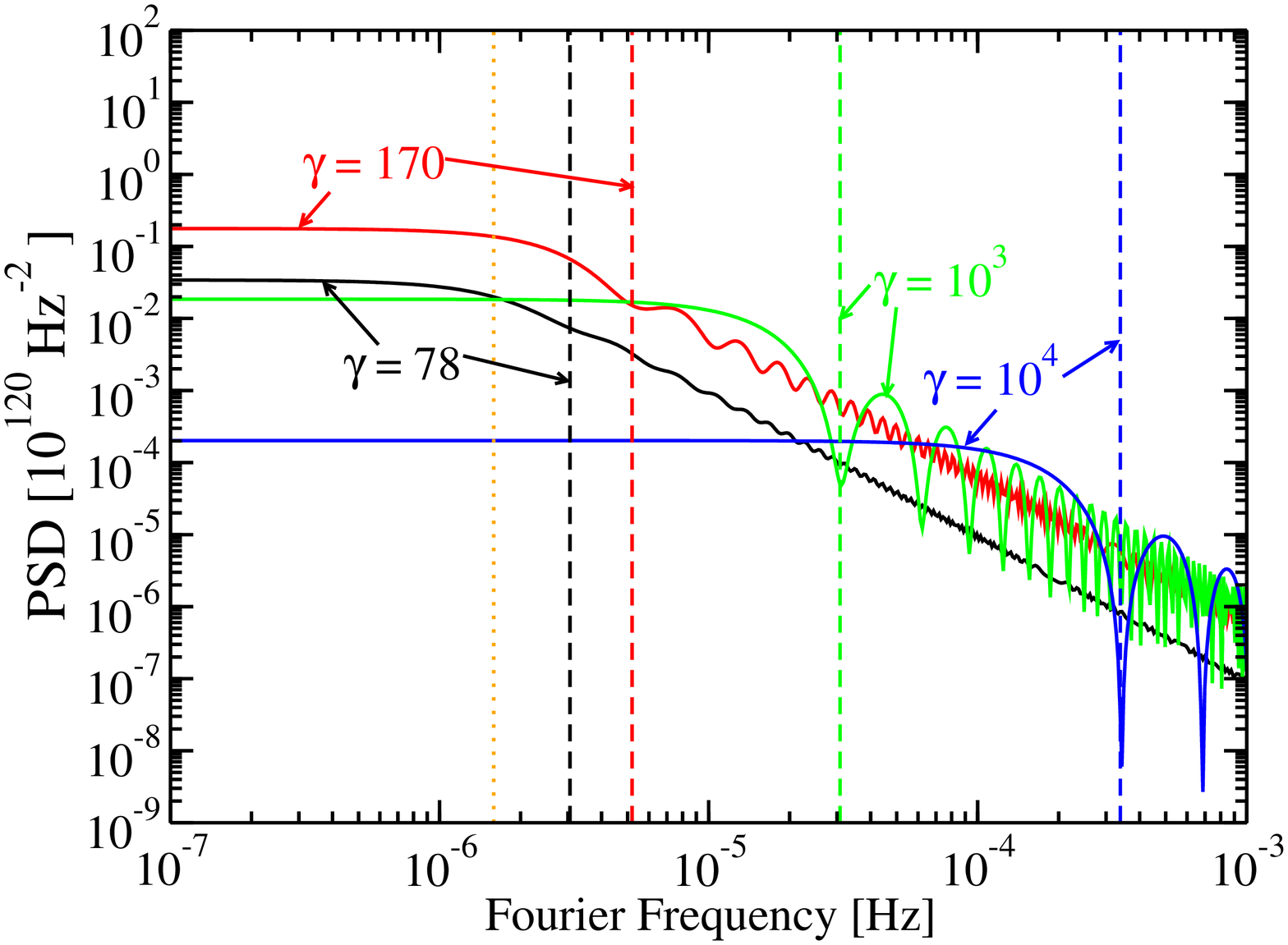}
\caption{The electron PSD from Equation (\ref{PSD_eqn2})
resulting from an instantaneous flash ($a=0$) of electrons injected
with a power-law energy index $q=2$.  Here we set $t_{\rm esc}=10^5\
\s$, $\nu=3.1\times10^{-8}\ \s^{-1}$, $\langle L_{\rm
inj}\rangle=10^{42}\ \erg\ \s^{-1}$, $\Delta t=1$\ year, $\g_1=10^2$,
$\g_2=10^5$.  Dashed lines indicate $f=t_{\rm cool}^{-1}$ for each
curve, and the dotted line indicates $f=(2\pi t_{\rm esc})^{-1}$.  }
\label{PSDfig}
\vspace{2.2mm}
\end{figure}

We identify asymptotes for the PSD for $q=2$, Equation
(\ref{PSD_eqn2}).  For these asymptotes we assume $\g \ll \g_2$.

\begin{enumerate}
\item If $1/(\nu t_{\rm esc}) \ll \g$, and $2\pi f/\nu \ll \g$ then
\begin{eqnarray}
\label{asy1}
S(\g,f) \approx \frac{Q_0^2 (f/f_0)^{-a}}{\nu^2\g^2} \ .
\end{eqnarray}

\item If $1/(\nu t_{\rm esc}) \ll \g \ll 2\pi f/\nu$ then
\begin{eqnarray}
\label{asy2}
S(\g,f) \approx \frac{ Q_0^2 (f/f_0)^{-a-2} }{ f_0^2 \pi^2}
\sin^2\left(\frac{\pi f}{\nu\g}\right)\ .
\end{eqnarray}

\item If $\g \ll 1/(\nu t_{\rm esc})$ then
\begin{flalign}
\label{asy3}
S(\g,f) & \approx \frac{Q_0^2 (f/f_0)^{-a}}{1/t_{\rm esc}^2 + (2\pi f)^2}
\exp\left[ \frac{-2}{t_{\rm esc}} \left( \frac{1}{\nu\g} - t_{\rm cool} \right) \right]
\\ \nonumber & \times
\left\{ 1 - \exp\left[-\frac{t_{\rm cool}}{t_{\rm esc}}\right]
\cos[2\pi f t_{\rm cool}]  \right\}\ 
\end{flalign}
where we define $t_{\rm cool}^{-1} = \nu \max(\g,\g_1)$.

\end{enumerate}

The electron PSD resulting from Equation (\ref{PSD_eqn2}) is
plotted in Figure \ref{PSDfig} for parameters described in the
caption, which are fairly standard ones for FSRQs.  We use $a=0$
here, which represents an instantaneous injection of power-law
particles at $t=0$, to more easily display the observable features, a
number of which are present.  For the $\g=78$, 170, and $10^3$
curves, where $\g \ll (\nu t_{\rm esc})^{-1}$), a break in the power
law from
$$
S(\g,f)\propto f^{-a}$$ 
to 
$$
S(\g,f)\propto f^{-(a+2)}
$$ 
is apparent, and the break frequency is at 
$$
f\approx(2\pi t_{\rm esc})^{-1}\ .  
$$
This is in agreement with asymptote 3 above.  In general, by examining
asymptote 3, it is clear that for low $\g$ a break in the PSD will be
found at a frequency of $f=(2\pi t_{\rm esc})^{-1}$.  Since the PSD
measures periodic variability, this indicates that for low $\g$,
periodic variability on timescales less than the escape timescale is
less preferred.  This is because electrons will always be escaping at
a single timescale, which does not vary with time.  One can also see
in the $\g=170$ curve in Figure \ref{PSDfig} at high $f$ structure
related to the cosine seen in asymptote 3, with local minima at
integer multiples of $t_{\rm cool}$.

In the $\g=10^3$ and $\g=10^4$ curves, at high $\g$ ($\g \gg (\nu
t_{\rm esc})^{-1}$) the PSD will transition from 
$$
S(\g,f)\propto f^{-a}
$$ 
to
$$
S(\g,f)\propto f^{-(a+2)}\ ,
$$
but in this case the transition is at 
$$
f=t_{\rm cool}^{-1}\ ,
$$
in agreement with asymptotes 1 and 2.  Thus, variability on timescales
less than the cooling timescales will be less periodic, since cooling
on those smaller timescales will always be present.  It is also clear
that at high $f$ minima from the $\sin^2$ term in asymptote 2 occur at
integer multiples of $t_{\rm cool}^{-1}$.  Since at high values of
$\g$, where the cooling is strongest, cooling on timescales $t_{\rm
cool}$ will be immediate, and so periodic variability on timescales
that are integer multiples of $t_{\rm cool}$ will also be strongly
avoided.

\subsection{Electron Injection Index $q\ne2$}
\label{generalsoln}

\begin{figure}
\vspace{5.2mm} 
\epsscale{1.0} 
\plotone{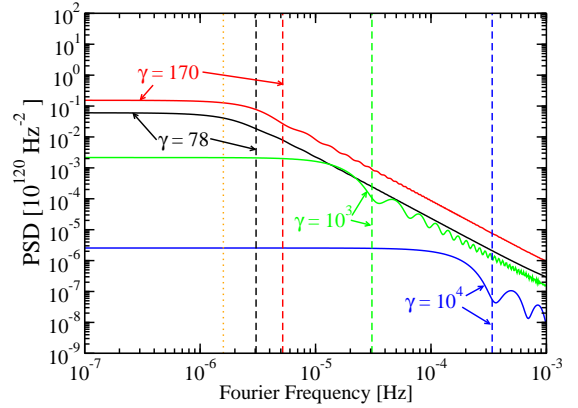}
\caption{The same as Fig.\ \ref{PSDfig}, except that we set the 
injection power-law index $q=2.5$.
}
\label{PSDq25fig}
\vspace{2.2mm}
\end{figure}

In this case, there is no simple analytic solution to equation
(\ref{Nsoln3}), although it can be written with the incomplete Gamma
function as
\begin{flalign}
\g^2\tilN_e(\g,f) & = Q_0 (f/f_0)^{-a/2} 
\exp\left[ \frac{-1}{\nu\g}\term \right] 
\nonumber \\ & \times
\nu^{q-2}\left(i\omega-\frac{1}{t_{\rm esc}}\right)^{1-q}
\nonumber \\ & \times 
\left[ \g_g(q-1,-u_{\max}) - \g_g(q-1,-u_{\min}) \right]\ .
\end{flalign}
We compute the function numerically, and results can be seen in Figure
\ref{PSDq25fig} for $q=2.5$.  This confirms the features seen in the
$q=2$ case are also seen for other values of $q$, although the minima
at high $\g$ are not as pronounced.

\section{Time Lags}
\label{lagsection}

The phase lag between two electron Lorentz factor ``channels'', $\g_a$ and $\g_b$ 
as a function of Fourier frequency ($f$) can be calculated from
\begin{eqnarray}
\Delta\phi(\g_a,\g_b,f) = \arctan\left\{ \frac{Y_I(\g_a,\g_b,f)}{Y_R(\g_a,\g_b,f)]} \right\}  \ 
\end{eqnarray}
where
\begin{eqnarray}
[\g_a^2 \tilN_e(\g_a,f)][\g_b^2 \tilN_e(\g_b,f)]^* = 
\nonumber \\
Y_R(\g_a,\g_b,f) + i Y_I(\g_a,\g_b,f)\ .
\end{eqnarray}
This implies that 
\begin{eqnarray}
Y_R(\g_a,\g_b,f) = Re[\g_a^2 \tilN_e(\g_a, f)]Re[\g_b^2 \tilN_e(\g_b, f)] 
\nonumber \\ 
+ Im[\g_a^2 \tilN_e(\g_a, f)]Im[\g_b^2 \tilN_e(\g_b, f)]
\end{eqnarray}
and 
\begin{eqnarray}
Y_I(\g_a,\g_b,f) =  Re[\g_b^2 \tilN_e(\g_b, f)]Im[\g_a^2 \tilN_e(\g_a, f)] 
\nonumber \\ 
- Re[\g_a^2 \tilN_e(\g_a, f)]Im[\g_b^2 \tilN_e(\g_b, f)] 
\ .
\end{eqnarray}
The time lag can be calculated from the phase lag, 
\begin{eqnarray}
\Delta T(\g_a, \g_b, f) = \frac{\Delta\phi(\g_a,\g_b,f)}{2\pi f} \ .
\end{eqnarray}

\begin{figure}
\vspace{5.2mm} 
\epsscale{1.0} 
\plotone{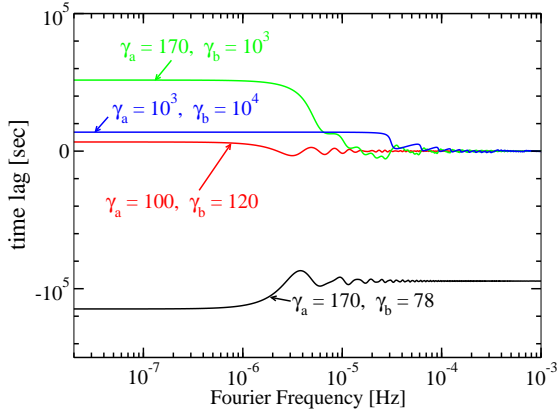}
\caption{The electron time lags with various values of $\g_a$ and $\g_b$.
Parameters are the same as in Figure \ref{PSDfig}.  }
\label{lagfig1}
\vspace{2.2mm}
\end{figure}

For our solution, Equation (\ref{Nsoln4}) the time lag is
\begin{eqnarray}
\Delta T(\g_a, \g_b, f) = \frac{1}{2\pi f} 
\arctan\left\{ \frac{Z_I(\g_a,\g_b,f)}{Z_R(\g_a,\g_b,f)}\right\}
\end{eqnarray}
where
\begin{flalign}
\label{ZI}
Z_{I}(\g_a,\g_b,f) & = \exp\left[-\frac{1}{\nu t_{\rm esc}}
\left(\frac{1}{\g_a} + \frac{1}{\g_b}\right)\right]
\nonumber \\ & \times
\Biggr\{ \exp\left[\frac{2}{\nu t_{\rm esc}\g_2}\right]
\sin\left[\frac{\omega}{\nu}\left(\frac{1}{\g_a}-\frac{1}{\g_b}\right)\right]
\nonumber \\ &
+ \exp\left[\frac{1}{\nu t_{\rm esc}}\left(\frac{1}{\max(\g_1,\g_a)} + 
\frac{1}{\max(\g_1,\g_b)}  \right)\right]
\nonumber \\ & \times
\sin\Biggr[\frac{\omega}{\nu}\Biggr(\frac{1}{\max(\g_1,\g_b)}-\frac{1}{\g_b} 
\nonumber \\ &
+ \frac{1}{\g_a} - \frac{1}{\max(\g_1,\g_b)}\Biggr)\Biggr]
\nonumber \\ &
- \exp\left[\frac{1}{\nu t_{\rm esc}}\left(\frac{1}{\max(\g_1,\g_a)} + 
\frac{1}{\g_2} \right)\right]
\nonumber \\ & \times
\sin\left[\frac{\omega}{\nu}\left(\frac{1}{\g_a}-\frac{1}{\max(\g_1,\g_a)} +
\frac{1}{\g_2} - \frac{1}{\g_b}\right)\right]
\nonumber \\ &
- \exp\left[\frac{1}{\nu t_{\rm esc}}\left(\frac{1}{\g_2}  + 
\frac{1}{\max(\g_1,\g_b)}  \right)\right]
\nonumber \\ & \times
\sin\left[\frac{\omega}{\nu}\left(\frac{1}{\g_a}-\frac{1}{\g_2} +
\frac{1}{\max(\g_1,\g_b)} - \frac{1}{\g_b} \right)\right]\ \Biggr\}
\end{flalign}

and $Z_R(\g_a,\g_b,f)$ is the same as $Z_I(\g_a,\g_b,f)$ except with
$\cos$ in place of $\sin$.  Several examples of electron time
lags can be seen in Fig.\ \ref{lagfig1}.

For $f\ll \nu\g_a/(2\pi)$ and $f\ll \nu\g_b/(2\pi)$ 
the time delay will be approximately independent of frequency, with value
\begin{eqnarray}
\Delta T(\g_a, \g_b, f) \approx \frac{1}{\nu} \frac{A_I(\g_a,\g_b)}{A_R(\g_a,\g_b)}
\end{eqnarray}
where
\begin{flalign}
A_I(\g_a,\g_b,f) & = 
\left(\frac{1}{\g_a}-\frac{1}{\g_b}\right) 
\exp\left[\frac{-1}{\nu t_{\rm esc}}\left(\frac{1}{\g_a}+\frac{1}{\g_b}\right)\right]
\nonumber \\ &
+ \frac{1}{\g_b} \exp\left[\frac{-1}{\nu t_{\rm esc}\g_b}\right]
\nonumber \\ &
- \frac{1}{\g_a} \exp\left[\frac{-1}{\nu t_{\rm esc}\g_a}\right]
\end{flalign}
and
\begin{flalign}
A_R(\g_a,\g_b,f) & = 1 + 
\exp\left[\frac{-1}{\nu t_{\rm esc}}\left(\frac{1}{\g_a}+\frac{1}{\g_b}\right)\right]
\nonumber \\ &
- \exp\left[\frac{-1}{\nu t_{\rm esc}\g_b}\right] 
- \exp\left[\frac{-1}{\nu t_{\rm esc}\g_a}\right]\ ,
\end{flalign}
where we also assumed $\g_1<\g_a\ll \g_2$ and $\g_1<\g_b\ll\g_2$.  At
these low values of $f$, the lags are positive for $\g_a>\g_b$
indicating the smaller $\g$ lags behind the larger $\g$.  This is due
to the fact that electrons with smaller $\g$ will take longer to cool
than those with larger $\g$.  If also $(\nu t_{\rm esc})^{-1}\ll \g_a$ and
$(\nu t_{\rm esc})^{-1}\ll \g_b$ then
\begin{eqnarray}
\label{lag_lowfreq}
\Delta T(\g_a,\g_b,f) \approx \frac{1}{2\nu}\left( \frac{1}{\g_a} - \frac{1}{\g_b}\right)\ .
\end{eqnarray}
An example of this can be seen in Fig.\ \ref{lagfig1}, with the
$\g_a=10^3$, $\g_b=10^4$ curve.  If $\g_a\ll (\nu t_{\rm esc})^{-1} \ll
\g_b$ then
\begin{eqnarray}
\Delta T(\g_a,\g_b,f) \approx t_{\rm esc}
\Biggr\{ 1 - \frac{1}{\nu t_{\rm esc}\g_b} 
\nonumber \\
- \frac{1}{\nu t_{\rm esc}\g_a} \exp\left[\frac{-1}{\nu t_{\rm esc}\g_a}\right] 
\Biggr\} \ .
\end{eqnarray}
In Fig.\ \ref{lagfig1}, an example can be seen with the $\g_a=170$,
$\g_b=10^3$ curve.

If $\g_a \ll (\nu t_{\rm esc})^{-1}$ and $\g_b\ll (\nu t_{\rm esc})^{-1}$ then
$\Delta T(\g_a,\g_b,f) \rightarrow 0$.  In Fig.\ \ref{lagfig1}, the
curve that most closely approximates this is the $\g_a=100$, $\g_b=120$
curve.

At high $f$, The behavior is quite complex.  By inspecting Equation
(\ref{ZI}), we can see that the important frequencies are those that
make the sine or cosine terms go to 0.  They will be the integer or
half-integer multiples of
\begin{eqnarray}
f=t_{{\rm cool},a}^{-1} = \nu\g_a
\\
f=t_{{\rm cool},b}^{-1} = \nu\g_b
\\ 
f=(t_{{\rm cool},a}-t_{{\rm cool},b})^{-1}\ .
\end{eqnarray}
There are many local minima and maxima at the integer or 
half-integer multiples of these values.

\section{Emission and Light Travel Time Effects}
\label{emission}

In the previous sections we have explored the PSD and time lags for
the electron distribution.  However, what is observed is the emission
of these electrons, through synchrotron or Compton-scattering.  We
will assume that the emitting region is spherical with co-moving
radius $R^\prime$ and homogeneous, containing a tangled magnetic field
of strength $B$.  The blob is moving with a relativistic speed $\beta
c$ ($c$ being the speed of light), giving it a bulk Lorentz factor
$\G=(1-\beta^2)^{-1/2}$.  The blob is moving with an angle $\theta$ to
the line of sight giving it a Doppler factor
$\delta_D=[\G(1-\beta\cos(\theta))]^{-1}$.  Although we make the
simplifying assumption the blob is homogeneous, with variations in the
electron distribution taking place throughout the blob simultaneously,
since it has a finite size, photons will reach the observer earlier
from the closer part than the farther part, and thus one must
integrate over the time in the past, $t^\prime$.  This is similar to
the ``time slices'' of \citet{chiaberge99}.  We note again that
Fourier frequencies higher than the inverse of the light crossing
timescale are unphysical in the simple model we present here.

\subsection{Synchrotron and External Compton}
\label{synch_section}

Taking the light travel time into account, in the
$\delta$-approximation the observed $\nu F_\nu$ flux at observed
energy $\e$ (in units of the electron rest energy) from synchrotron or
Compton scattering of an external isotropic monochromatic radiation
field as a function of the observer's time $t$ is
\begin{eqnarray}
\label{Fsy1}
F_\e(t) = \frac{K(1+z)}{t_{lc}\dD} \int^{2R^\prime/c}_0 d\tp\ 
N_e\left(\gp; \frac{t\delta_D}{1+z}-\tp\right)
\end{eqnarray}
where
\begin{eqnarray}
t_{lc} = \frac{2R^\prime (1+z)}{c \dD}\ 
\end{eqnarray}
is the light crossing time in the observer's frame.  A derivation of
the light travel time effect can be found in Appendix
\ref{lighttravelappendix}.  For synchrotron emission,
\begin{eqnarray}
K = K_{sy} = \frac{\delta_D^4}{6\pi d_L^2} c \sT u_B \gamma_{sy}^{\prime 3}\ ,
\end{eqnarray}
and
\begin{eqnarray}
\label{gamma_sy}
\gp = \gp_{sy} = \sqrt{\frac{\e(1+z)}{\delta_D\e_B}}\ .
\end{eqnarray}
The Thomson cross-section is $\sT=6.65\times10^{-25}\ \cm^2$, the
Poynting flux energy density is $u_B=B^2/(8\pi)$, $\e_B=B/B_{c}$ where
$B_c=4.414\times10^{13}\ \Gauss$, the redshift of the source is $z$,
and the luminosity distance to the source is $d_L$.  For external
Compton (EC) scattering,
\begin{eqnarray}
K = K_{EC} = \frac{\delta_D^6}{6\pi d_L^2} c \sT u_0 \gamma_{EC}^{\prime 3}\ ,
\end{eqnarray}
\begin{eqnarray}
\label{gamma_EC}
\gp = \gp_{EC} = \frac{1}{\delta_D}\sqrt{\frac{\e(1+z)}{2 \e_0}}\ 
\end{eqnarray}
\citep{dermer09_book}.  Here the external radiation field energy
density and photon energy (in units of the electron rest energy) are
$u_0$ and $\e_0$, respectively.  This approximation is valid in the
Thomson regime, i.e., when $\gp \la (\G\e_0)^{-1}$.  Primed quantities
refer to the frame co-moving with the emitting region.  The cooling 
rate parameter is
\begin{eqnarray}
\label{nu}
\nu=\frac{4}{3 m_e c^2}c \sT (u_B+\G^2 u_0)\ 
\end{eqnarray}
where we ignore the effects of SSC cooling.  

It is shown in Appendix \ref{lighttravelfourierappendix} that the 
Fourier transform of Equation (\ref{Fsy1}) is
\begin{eqnarray}
\label{tilF5}
\tilF_\e(f) = \frac{K(1+z)}{2\pi if t_{lc}\dD} \tilN_e\left(\gp, \frac{(1+z)f}{\dD}\right)\ 
\nonumber \\
\left\{ \exp\left[ \frac{4 \pi i f(1+z) R^\prime}{c\dD}\right] - 1 \right\}\ .
\end{eqnarray}

Equation (\ref{tilF5}) implies the PSD of the synchrotron or EC
flux is
\begin{flalign}
\label{PSDflux_eqn}
S(\e,f) & = |\tilF_\e(f)|^2 
\nonumber \\ &
= \frac{K^2(1+z)^2}{(\pi f t_{lc}\dD)^2} \left|\tilN_e\left(\gp, \frac{(1+z)f}{\dD}\right)\right|^2
\sin^2\left( \pi f t_{lc} \right)\ .
\end{flalign}
If the emitting region is very compact, i.e., if $R^\prime \ll c\dD(2f(1+z))^{-1}$, then
\begin{eqnarray}
\label{PSDflux_eqn2}
S(\e,f) \approx \frac{K^2(1+z)^2}{(\pi f t_{lc}\dD)^2} \left|\tilN_e\left(\gp, \frac{(1+z)f}{\dD}\right)\right|^2
\left( \pi f t_{lc} \right)^2
\nonumber \\
S(\e,f) \approx \frac{K^2(1+z)^2}{\dD^2} \left|\tilN_e\left(\gp, \frac{(1+z)f}{\dD}\right)\right|^2\ .
\end{eqnarray}
So in the case of a very compact emitting region, the light travel
time effects play no part in the PSD, as one would expect.

\begin{figure}
\vspace{5.2mm} 
\epsscale{1.0} 
\plotone{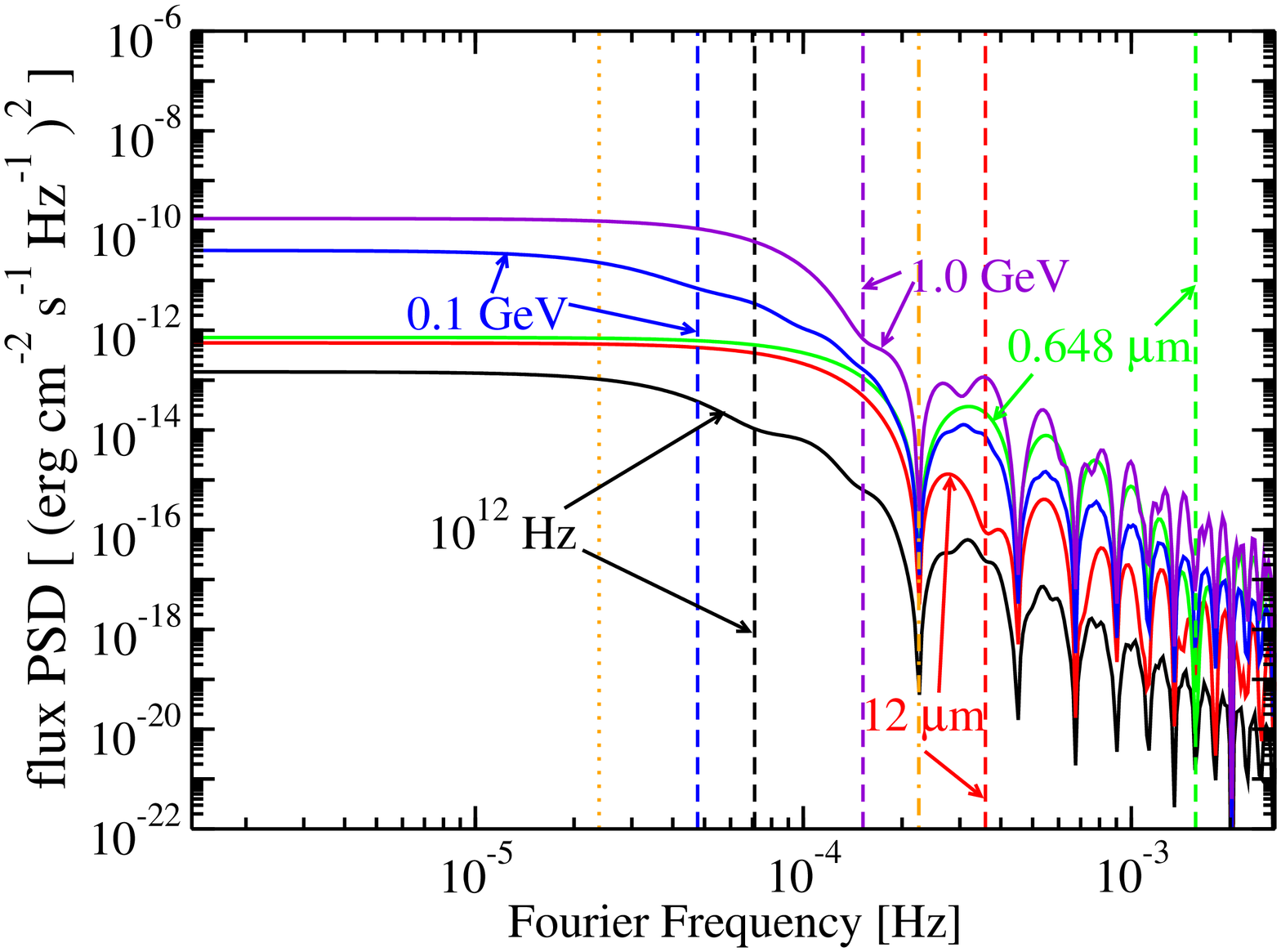}
\caption{The flux PSD computed from Equations (\ref{PSDflux_eqn}) and
(\ref{PSD_eqn2}) using the same parameters as in Figure \ref{PSDfig}.
Additional parameters are $\delta_D=\Gamma=30$, $B=1$\ G,
$u_0=10^{-3}\ \erg\ \cm^{-3}$, $\e_0=2\times10^{-5}$,
$R^\prime=10^{15}\ \cm$, and $z=1$.  At this redshift with a cosmology
$(h, \Omega_m, \Omega_\Lambda)= (0.7, 0.3, 0.7)$, $d_L=2\times10^{28}\
\cm$.  The observed photon frequency, wavelength, or energy is shown,
along with $t_{\rm cool}^{-1}$ for each curve (dashed lines), $(2\pi
t_{\rm esc})^{-1}$ (dotted line), and $t_{lc}^{-1}$ (dashed-dotted
line), all computed in the observer's frame.  }
\label{PSD_flux_fig}
\vspace{2.2mm}
\end{figure}

The observed flux PSDs from synchrotron and EC one would expect are
shown in Figure \ref{PSD_flux_fig}, calculated from Equations
(\ref{PSDflux_eqn}) and (\ref{PSD_eqn2}).   The parameter values 
are the same as in Figure \ref{PSDfig}, with additional parameters given in the
caption.  The seed photon source is assumed to be Ly$\alpha$ photons,
presumably from a broad-line region.  Again, parameters are chosen to
be consistent with those one would expect from an FSRQ.  Synchrotron
PSDs are shown for 12 Hz, 12 $\mu$m (the {\em WISE} W3 filter's
central wavelength), and 0.648 $\mu$m (the central wavelength of the
Johnson $R$ band).  EC PSDs are shown for 0.1 and 1.0 GeV, which are
within the \fermi-LAT energy range.  Frequencies lower than $10^{12}$
Hz can be affected by synchrotron self-absorption, which is not
considered in this paper.  X-rays are not shown as they are likely
dominated by SSC emission, which is considered in Section
\ref{SSC_section} below.  Also note that in real FSRQs the 12 $\mu$m
PSD could suffer from contamination from dust torus emission
\citep[e.g.,][]{malmrose11} and the $R$ band could suffer from
contamination from accretion disk emission, and we do not take either
of these possibilities into account.

Figure \ref{PSD_flux_fig} shows many of the features seen in Figure
\ref{PSDfig}.  For photons generated from electrons with low $\gp$ the
PSDs show a break from 
$$
S(\e,f)\propto f^{-a}
$$
to 
$$
S(\e,f)\propto f^{-(a+2)}
$$
at approximately
$$
f=(2\pi t_{\rm esc})^{-1}\ ,
$$
as seen in the $10^{12}$ Hz and 0.1 GeV PSDs.  For synchrotron, if
$u_B\ll \G^2u_{0}$, as is usually the case for FSRQs, this regime
occurs when
\begin{flalign}
\nu & \ll \nu_{\rm cr,sy} = 10^{13}\ \Hz\ \left(\frac{\delta_D}{\G}\right) 
\left(\frac{\G}{30} \right)^{-3}
\nonumber \\ & \times
\left(\frac{u_{0}}{10^{-3}\ \erg\ \cm^{-3}} \right)^{-2}
\left(\frac{t_{\rm esc}}{10^5\ \s} \right)^{-2}
\left(\frac{B}{1\ \Gauss} \right)
\frac{1}{1+z}\ .
\end{flalign}
For EC, this regime occurs when
\begin{flalign}
m_e c^2\e & \ll E_{\rm cr,EC}  = 
2\ \GeV \left(\frac{\delta_D}{\G} \right)^{2}
\left(\frac{\G}{30} \right)^{-2}
\nonumber \\ & \times
\left(\frac{u_{0}}{10^{-3}\ \erg\ \cm^{-3}} \right)^{-2}
\left(\frac{t_{\rm esc}}{10^5} \right)^{-2}
\nonumber \\ & \times
\left(\frac{\e_0}{2\times10^{-5}} \right)
\frac{1}{1+z}\ .
\end{flalign}
If $u_B \gg \G^2 u_{0}$ then
\begin{flalign}
\nu_{\rm cr,sy} & = 5\times10^{15}\ \Hz \left(\frac{\delta_D}{30}\right)^2 
\left(\frac{t_{\rm esc}}{10^5\ \s} \right)^{-2}
\nonumber \\ & \times
\left(\frac{B}{1\ \Gauss} \right)^{-3}
\frac{1}{1+z}\ 
\end{flalign}
and
\begin{flalign}
E_{\rm cr,EC} & = 1\ \TeV\ \left(\frac{\delta_D}{30} \right)^{2}
\left(\frac{t_{\rm esc}}{10^5} \right)^{-2}
\left(\frac{B}{1\ \Gauss} \right)^{-4}
\nonumber \\ & \times
\left(\frac{\e_0}{2\times10^{-5}} \right)
\frac{1}{1+z}\ .
\end{flalign}

For photons generated from electrons with high $\gp$, ($\nu\gg
\nu_{\rm cr,sy}$ or $m_ec^2\e \gg E_{\rm cr,EC}$ for synchrotron or
EC, respectively) minima are seen at integer multiples of $f=t_{\rm
cool}^{-1}$, as seen in the 12 $\mu$m, $R$ band and 1.0 GeV PSDs, as
well as a break of 2 at approximately $f=t_{\rm cool}^{-1}$.  There is
an additional feature in Figure \ref{PSD_flux_fig} not seen in Figure
\ref{PSDfig} related to the light travel timescale.  In agreement with
Equation (\ref{PSDflux_eqn}), $\sin^2$ minima can be seen at integer
multiples of $f=1/t_{lc}$ which appear in the PSDs at all photon
energies.  Additionally, there is a break of 2 at
approximately this frequency, in agreement with Equations
(\ref{PSDflux_eqn}) and (\ref{PSDflux_eqn2}).

\begin{figure}
\vspace{2.2mm} 
\epsscale{1.0} 
\plotone{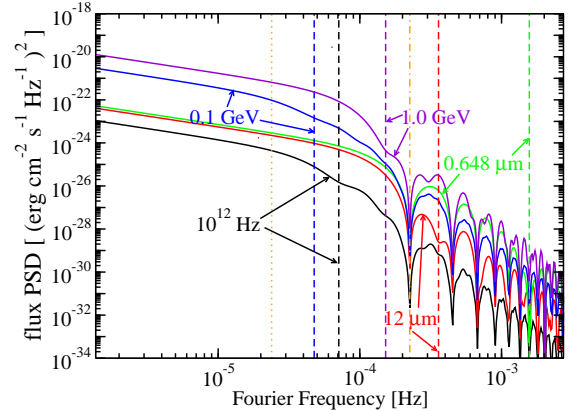}
\caption{The same as Fig.\ \ref{PSD_flux_fig} except
that we set $a=1$.}
\label{PSD_flux_fig_a1}
\vspace{2.2mm}
\end{figure}

In Figure \ref{PSD_flux_fig_a1} we plot observed PSDs for the
parameters $a=1$.  This demonstrates that our model can reproduce any
color noise in a synchrotron or EC PSD for an appropriate choice of
$a$, and that the features described above are preserved for different
values of $a$.

\begin{figure}
\vspace{2.2mm} 
\epsscale{1.0} 
\plotone{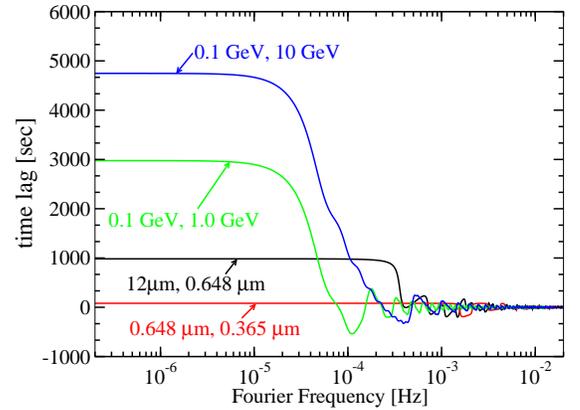}
\caption{The observed time lags as a function of observed Fourier
frequency.  Parameters are the same as in Figure \ref{PSDfig}, with
parameters the same as Figures \ref{PSDfig} and \ref{PSD_flux_fig}.
The 0.648 $\mu$m (central wavelength of $R$ band) and 0.365 $\mu$m
(central wavelength of $U$ band) are from synchrotron, while the 0.1,
1.0, and 10 GeV emission is from EC.  }
\label{lagfig2}
\vspace{2.2mm}
\end{figure}

For synchrotron or EC emission, in our simple model, the
time delays will play no role in the Fourier frequency-dependent time
lags (Section \ref{lagsection}).  It is easy enough to see by
inspecting Equation (\ref{tilF5}) that terms associated with the
light travel time will cancel when calculating time lags.  However, 
one must be careful to shift the time lag and frequency into the 
observed frame by multiplying by $(1+z)/\dD$ and $\dD/(1+z)$, 
respectively.  An example of observed time lags can be 
seen in Fig.\ \ref{lagfig2} for synchrotron emission and 
EC emission.  

\subsection{Synchrotron Self-Compton}
\label{SSC_section}

The synchrotron-producing electrons will also Compton scatter the
synchrotron radiation they produce, leading to SSC emission.  Again,
assuming the blob is homogeneous, and taking into account light travel
time effects, we have
\begin{flalign}
F_\e^{SSC}(t) & = \frac{K_{SSC}(1+z)}{t_{lc}\delta_D}
\int^{2R^\prime/c}_0 dt^\prime\ 
\int_0^{\min[\ep,\e^{\prime -1}]}
\frac{d\ep_i}{\ep_i}\ 
\nonumber \\ & \times 
N_e\left(\sqrt{\frac{\ep}{\ep_i}}; 
\frac{t\dD}{1+z}-t^\prime \right)\ 
N_e\left( \sqrt{\frac{\ep_i}{\e_B}}; 
\frac{t\dD}{1+z}-t^\prime  \right)
\end{flalign}
where
\begin{eqnarray}
K_{SSC} = \frac{ \delta_D^4 c\sT^2 R^\prime u_B}{12\pi d_L^2 V^\prime}
\left(\frac{\e}{\e_B}\right)^{3/2}\ ,
\end{eqnarray}
\begin{eqnarray}
V^\prime = \frac{4}{3}\pi R^{\prime 3}\ 
\end{eqnarray}
is the blob volume in the comoving frame, and
\begin{eqnarray}
\ep = \frac{(1+z)\e}{\delta_D}\ 
\end{eqnarray}
\citep{dermer09_book}.  Following the same procedure for 
synchrotron and EC, for the Fourier transform we get
\begin{flalign}
\tilF_\e^{SSC}(t) & = \frac{K_{SSC}(1+z)}{2\pi i t_{lc} f \dD}
\left\{ \exp\left[ \frac{4 \pi i f(1+z) R^\prime}{c\dD}\right] - 1 \right\}
\nonumber \\  & \times
\int_{-\infty}^{\infty}df^\prime\ 
\int_0^{\min[\ep,\e^{\prime -1}]}
\frac{d\ep_i}{\ep_i}\ 
\nonumber \\ & \times
\tilN_e\left( \sqrt{\frac{\ep}{\ep_i}};
\frac{(1+z)f}{\dD}-f^\prime  \right)
\tilN_e\left( \sqrt{\frac{\ep_i}{\e_B}}; 
f^\prime \right)\ .
\end{flalign}

Note, however, that we ignore the effects of SSC cooling.  In this
case, $\dotg(\g)$ would be dependent on the electron distribution,
$\g^2N(\g;t)$ (or $\g^2\tilN(\g,f)$) leading to a non-linear
differential equation.  This is treated in detail for the continuity
equation by \citet{schlickeiser09, schlickeiser10}; and 
\citet{zacharias10,zacharias12, zacharias12_EC}.  The SSC PSD is
\begin{eqnarray}
\label{SSC_PSD_eqn}
S^{SSC}(\e,f) = | \tilF_\e^{SSC}(t) |^2 =
\nonumber \\
\left( \frac{K_{SSC}(1+z)}{2\pi t_{lc} f \dD}\right)^2 
\sin^2(\pi f t_{lc}) |I(\e,f)|^2
\end{eqnarray}
where
\begin{flalign}
I(\e,f) & = \int_{-\infty}^{\infty}df^\prime\ 
\int_0^{\min[\ep,\e^{\prime -1}]}
\frac{d\ep_i}{\ep_i}\ 
\nonumber \\  & \times
\tilN_e\left( \sqrt{\frac{\ep}{\ep_i}};
\frac{(1+z)f}{\dD}-f^\prime  \right)
\tilN_e\left( \sqrt{\frac{\ep_i}{\e_B}}; 
f^\prime \right)\ .
\end{flalign}

\begin{figure}
\vspace{5.2mm} 
\epsscale{1.0} 
\plotone{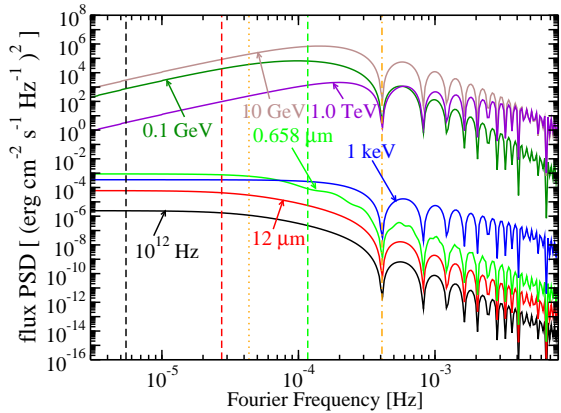}
\caption{The synchrotron and SSC flux PSD computed from Equation
(\ref{SSC_PSD_eqn}).  Parameters are the same as in Figure
\ref{PSD_flux_fig}, except $u_0=0$ and $z=0.1$, giving
$d_L=1.4\times10^{27}\ \cm$ with a cosmology where $(h, \Omega_m,
\Omega_\Lambda)= (0.7, 0.3, 0.7)$.  For the synchrotron curves, the
frequency associated with $t_{\rm cool}$ for the electrons that
produce those photons is shown as the dashed lines.  The dotted curve
indicates the frequency $(2\pi t_{\rm esc})^{-1}$ and the
dashed-dotted line indicates the frequency $t_{lc}^{-1}$, all computed
in the observer's frame.  }
\label{PSD_SSC_fig}
\vspace{2.2mm}
\end{figure}

At low frequencies, $f \ll (\pi t_{lc})^{-1}$ and $f^\prime \ll (\pi
t^\prime_{\rm esc})^{-1}$,
\begin{flalign}
S^{SSC}(\e,f) & \approx \frac{K_{SSC}^2(1+z)^2}{4\dD^2} 
(Q_0 t_{\rm esc})^4 \left(\frac{\dD\e_B}{\e(1+z)}\right)^2
\nonumber \\ & \times
\left[ C_1(a, \e_{\min},\e_{\max})
C_2(a, f_{\min}, f_{\max}) \right]^2
\end{flalign}
where
\begin{flalign}
C_1(a, \e_{\min}, \e_{\max}) & = \int_{\e_{min}}^{\e_{max}} \frac{d\ep_i}{\ep_i}
\nonumber \\ & \times
\Biggr\{ \exp\left[ \frac{-1}{\nu t_{\rm esc}}\left( \sqrt{\frac{\ep_i}{\ep}} + 
  \sqrt{\frac{\e_b}{\ep_i}} \right)\right] 
\nonumber \\ & 
- \exp\left[\frac{-1}{\nu t_{\rm esc}} \sqrt{\frac{\ep_i}{\ep}} \right] 
\nonumber \\ & 
- \exp\left[\frac{-1}{\nu t_{\rm esc}}\sqrt{\frac{\e_b}{\e_i}}\right]  + 1 \Biggr\}\ ,
\end{flalign}
\begin{flalign}
C_2(a, f_{\min}, f_{\max})  & =  f_{0}^a 
\nonumber \\ & 
\times
\left\{ \begin{array}{ll} (f_{\max}^{1-a}-f_{\min}^{1-a})(1-a)^{-1} & a \ne 1 \\
\ln(f_{\max}/f_{\min}) & a = 1
\end{array}
\right. \ ,
\end{flalign}
\begin{eqnarray}
f_{\min} = \max\left[ f_1, \frac{(1+z)f}{\dD}-f_2 \right]\ ,
\end{eqnarray}
\begin{eqnarray}
f_{\max} = \min\left[ f_2, \frac{(1+z)f}{\dD} - f_1 \right]\ ,
\end{eqnarray}
\begin{eqnarray}
\e_{\min} = \max[ \g_1^2\e_B, \ep/\g_2 ]\ ,
\end{eqnarray}
and
\begin{eqnarray}
\e_{\max} = \min[ \ep, \epsilon^{\prime -1}, \g_2^2\e_B, \ep/\g_1^2 ]\ .
\end{eqnarray}
If $f_2>(1+z)f/\dD \gg f_1$, then at low frequencies the
SSC PSD will go as 
$$
S^{SSC}(\e,f)\propto f^{-(2a-2)}\ .
$$
Synchrotron and SSC PSDs are shown in Figure
\ref{PSD_SSC_fig}.  The parameters were chosen to be those one would
expect from a high-peaked BL Lac object.  At low frequency, they agree
with the asymptote above.  Here, the PSDs are flatter (the PSD
power-law index is smaller) than for synchrotron or EC.  At high
frequency the behavior is complex, but the features from the light
crossing timescale are apparent.  There are no features associated
with the cooling timescale of the electrons which produce SSC photons,
since the SSC emission for a particular observed energy will be
produced by a broad range of electrons.  This is in contrast to
synchrotron or EC where the photons at a particular energy can be
approximated as being produced by a single electron Lorentz factor.

\section{Applications}
\label{applications}

The PSDs of blazars are almost always power-laws,
$S(\e,f)\propto f^{-b}$ (i.e., colored noise), although there is
sometimes evidence that they deviate from this.  As seen in Section
\ref{emission}, our model can reproduce this, since in our model at
low frequency for synchrotron or EC the PSD goes as
$S^{sy/EC}(\e,f)\propto f^{-a}$ and for SSC $S^{SSC}(\e,f)\propto
f^{-(2a-2)}$, where $a$ is the parameter from Equation (\ref{tilQ}).  
For an appropriate choice of $a$, it can reproduce any
power-law PSD.  At higher frequencies, our model predicts features in
PSDs that deviate from a strict single power-law.  In this section, we
explore some of the applications of our model to observed PSDs from
the literature.

\subsection{The VHE Gamma-ray PSD of PKS 2155$-$304}
\label{VHEPSD}

In the PSD measured by HESS from \object{PKS 2155$-$304},
\citet{aharonian07_2155} found a power law with $S(\e,f) \propto
f^{-2}$, out to $f\ga 10^{-3}$\ Hz.  There does appear to be a minimum
feature at $f\approx1.2\times10^{-3}$\ Hz.  It is not clear if there
is a break in the PSD at higher frequencies that this.  If this
feature is associated with the light crossing timescale, then
\begin{eqnarray}
R^\prime \approx \frac{c t_{lc} \dD}{2(1+z)} = 6.5\times10^{14} \left( \frac{\dD}{60}\right)\ \cm\ .
\end{eqnarray}
Such a high Doppler factor is needed for these flares to avoid
$\gamma\gamma$ attenuation \citep{begelman08,finke08_SSC}.  Since the
$\g$-rays from this source are likely associated with the SSC
mechanism for this source, this association in unambiguous.  There are
no PSD features associated with a cooling timescale for SSC, as shown in 
Section \ref{SSC_section}.

\subsection{The X-ray Timing Properties of Mrk 421}

\citet{zhang02_mrk421} constructed PSDs from {\em BeppoSAX} data on 
\object{Mrk 421}.  They construct PSDs with data from two energy intervals:  
0.1--2 keV, and 2--10 keV (see their Figure 4).  Both PSDs show 
a minimum feature at $f\approx 4.3\times10^{-5}$\ Hz.  Although 
the feature is tentative, since it is only dependent on one point 
for each PSD, and there are no error bars, the fact that the 
feature is at the same frequency in both energy bands is a hint 
that the feature is associated with the light crossing timescale.  In 
this case, $t_{lc}=2.3\times10^4\ \s = 6.4$\ hrs and
\begin{eqnarray}
R^\prime \approx \frac{c t_{lc} \dD}{2(1+z)} = 1.0\times10^{16} \left( \frac{\dD}{30}\right)\ \cm\ .
\end{eqnarray}

\citet{zhang02_mrk421} also provides time lags as a function of Fourier
frequency between the two energy bands, 0.1--2 keV and 2--10 keV (see
their Figure 5), giving us an opportunity to compare them with the
results of Section \ref{lagsection}.  \citet{zhang02_mrk421} finds time
lags at $f\la 10^{-4}$\ Hz that are approximately constant at $\Delta
T \approx 10^3\ \s$.  The approximate independence of the lag with
frequency implies that the lag is in the regime where $(\nu
t_{\rm esc})^{-1}\ll \g_a$ and $(\nu t_{\rm esc})^{-1}\ll \g_b$, and the lag
$\Delta T^\prime$ can be approximated by Equation (\ref{lag_lowfreq}).  
Note that this equation gives the lag in the jet comoving frame; in the 
observer's frame, $\Delta T = \Delta T^\prime (1+z)/\delta_D$.  The X-rays 
for \object{Mrk 421} are likely produced by synchrotron emission and 
the external energy density ($u_0$) is likely to be negligible for a BL Lac object, 
so one can combine Equation (\ref{lag_lowfreq}) with Equations (\ref{gamma_sy}) 
and (\ref{nu}) to get 
\begin{eqnarray}
B = \frac{B_c}{(\delta_D\e_a \e_b)^{1/3}}
\left\{ \frac{3(1+z) m_e c^2 (\e_b^{1/2}-\e_a^{1/2})}
{8 c\sT u_{Bc} \Delta T} \right\}^{2/3} 
\end{eqnarray}
where $u_{Bc} = B_c^2/(8\pi)$.  
With $z=0.03$ for \object{Mrk 421} and $m_ec^2 \e_a = 0.1$\ keV and $m_ec^2\e_b = 2$\ keV, 
\begin{eqnarray}
B = 0.7\ \left( \frac{\dD}{30}\right)^{-1/3} 
\left( \frac{\Delta T}{10^3\ \s} \right)^{-2/3}   \Gauss\ .
\end{eqnarray}

\subsection{The Gamma-Ray PSDs of FSRQs and BL Lac Objects}

\citet{nakagawa13} used more than four years of {\em Fermi}-LAT
data to compute the PSD of 15 blazars.  Each PSD is fit with either a
single or broken power-law model.  Their values of $b$ from their
fits, where $S(\e,f)\propto f^{-b}$ from the single power-law fit or
lower index from the broken power-law fit if that fit is statistically
significant are given in Table \ref{PSD_nakagawa}.  We neglect the
FSRQ S4 1030+61, which has poor statistics.  There is a clear
separation between $b$ for FSRQs and BL Lac objects.  All BL Lac
objects have $b\le 0.6$, while all FSRQs except for PKS 1222+216 have
$b>0.7$.  PKS 1222+216 seems to be an outlier in terms of its PSD
power-law index, although its $b$ is still larger than for any of the
BL Lac objects.  We also compute the value of $a$ from
our model (recall its definition in Equation [\ref{tilQ}]) needed to
reproduce the values of $b$.  For FSRQs, presumably emitting by EC,
this is just
$$
a = b_{EC}\ ,
$$
while for BL Lac objects, presumably emitting by SSC, this is
$$
a = \frac{b_{SSC}+2}{2} \ 
$$
(recall Section \ref{SSC_section}).  The mean values of $a$ for FSRQs
and BL Lac objects are within one standard deviation (S.D.)  of each
other.  The values of $b$ from \citet{nakagawa13} are in agreement
with our theory with values of $a$ that cluster around $a\sim
1$.

\begin{deluxetable}{ccc}
\tabletypesize{\scriptsize}
\tablecaption{{\em Fermi}-LAT PSD power-law indices ($b$) from
\citet{nakagawa13} and the values of $a$ from our model needed to
reproduce them.}
\tablewidth{0pt}
\tablehead{
\colhead{Object} &
\colhead{$b$} &
\colhead{$a$}
}
\startdata
\multicolumn{3}{c}{FSRQs} \\
\hline
\object{4C +28.07} & $0.93\pm0.23$ & 0.93 \\
\object{PKS 0426$-$380}\footnotemark[1] & $1.16\pm0.47$ & 1.16 \\ 
\object{PKS 0454$-$234} & $0.78\pm0.27$ & 0.78 \\
\object{PKS 0537$-$441}\tablenotemark{1} & $0.86\pm0.64$ & 0.86 \\
\object{PKS 1222+216}   & $0.65\pm0.21$ & 0.65 \\ 
\object{3C~273} & $1.30\pm0.27$ & 1.30 \\
\object{3C~279} & $1.23\pm0.35$ & 1.23 \\
\object{PKS 1510$-$089} & $1.10\pm0.30$ & 1.10 \\
\object{3C~454.3} & $1.00\pm0.24$ & 1.00 \\
\object{PKS 2326$-$502} & $1.26\pm0.44$ & 1.26 \\
\hline
mean & 1.01 & $1.01$ \\
S.D. & 0.26 & 0.26 \\
\hline\hline
\multicolumn{3}{c}{BL Lac objects} \\
\hline
\object{3C~66A} & $0.60\pm0.44$ & 1.22\\
\object{Mrk 421} & $0.38\pm0.21$ & 1.19 \\
\object{PKS 2155$-$304} & $0.58\pm0.33$ & 1.29\\
\object{BL~Lac} & $0.41\pm0.47$ & 1.21\\
\hline
mean & 0.49 & $1.23$ \\
S.D. & 0.11 & 0.07 
\enddata
\tablenotetext{1}{PKS 0426$-$380 and PKS 0537$-$441 were  
previously classified as BL Lac objects.}
\label{PSD_nakagawa}
\end{deluxetable} 

Based on the traditional classification, PKS 0537$-$441 and PKS
0426$-$380 are classified as BL Lac objects.  However, based on the
new classification by \citet{ghisellini11_transition} they are
considered FSRQs \citep[see also][]{sbarrato12_transition}.  We use
the more recent classification.  For a discussion see
\citet{dammando13_0537} for PKS 0537$-$441 and \citet{tanaka13} for
PKS 0426$-$380.

The PSD indices found by \citet{nakagawa13} are in general not
reproduced by an independent analysis by \citet{sobolewska14}.  This
could be due to the different analysis techniques used by the
different authors.  More work is needed to resolve the discrepancies.

Ideally, to test our theory one would want simultaneous
light curves from synchrotron emission (say, optical) and GeV
$\gamma$-ray emission over a long timescale.  The PSDs could be
computed from these light curves.  One expects that if the
$\gamma$-rays are emitted by EC, they would have the same $b$ as
synchrotron ($b_{sy}=b_{EC}=a$).  But if the $\gamma$-rays are
produced by SSC emission, one would expect $b$ to be less steep than
for the $\gamma$-rays compared to the optical, with the relation
between the SSC and synchrotron PSD indices given by
$b_{SSC}=2b_{sy}-2=2a-2$.  Of course this could be complicated by
emission from a thermal accretion disk unrelated to the jet
emission. 

\subsection{The Gamma-Ray PSD of 3C~454.3}
\label{psd_3c454.3}

The PSD of 3C~454.3 from \citet{nakagawa13} shows a break at frequency
$f_{brk} = 1.5\times10^{-6}$\ Hz, corresponding to a timescale of
$6.8\times10^5\ \s = 190$\ hours = 7.9 days.  Their broken power-law
fit to the PSD, where $S(\e,f) \propto f^{-b_1}$ at $f<f_{brk}$ and
$S(\e,f) \propto f^{-b_2}$ at $f>f_{brk}$ shows $b_1\approx 1$ and
$b_2\approx 3$.  This is a break of about 2, which is what is
expected from our theory.  The timescale could correspond to the light
crossing, cooling, or escape timescales.  If one interprets it as the
light-crossing timescale, $t_{lc}$, then
\begin{eqnarray}
R^\prime \approx \frac{c t_{lc} \dD}{2(1+z)} = 1.7\times10^{17}\left( \frac{\dD}{30}\right)\ \cm\ .
\end{eqnarray}
Interpreting it as the cooling timescale, $t_{\rm cool}$, and assuming $\dD=\G$, 
the external radiation field is
\begin{flalign}
u_0 & \approx \frac{3m_e c^2}{4 c\sT \G^2 t_{\rm cool}^\prime \gp} 
\nonumber \\ &
= 9.6\times10^{-6} \left(\frac{\G}{30}\right)^{-2}
\left( \frac{E}{100\ \MeV}\right)^{-1/2}
\nonumber \\ &\times
\left( \frac{\e_0}{5\times10^{-7}}\right)^{1/2} \erg\ \cm^{-3}\ 
\nonumber \\ &
= 6.1\times10^{-5} \left(\frac{\G}{30}\right)^{-2}
\left( \frac{E}{100\ \MeV}\right)^{-1/2}
\nonumber \\ & \times
\left( \frac{\e_0}{2\times10^{-5}}\right)^{1/2} \erg\ \cm^{-3}\ 
\end{flalign}
where $E$ is the observed photon energy.  The first line assumes the
seed photon source is a dust torus with temperature 1000 K; the second
assumes it is Ly$\alpha$, presumably from the broad line region.  Both
numbers give rather low values for $u_0$.

However, interpreting the break as either the cooling timescale or the
light crossing timescale is problematic, since variations on
timescales much shorter than this have been observed from \object{3C
454.3}, including decreases on much faster timescales
\citep{ackermann10_3c454.3,abdo11_3c454.3}.  But one could also
associate the break with the escape timescale for electrons in the
blob, as shown in Section \ref{synch_section}.  This break will occur
at $f=(2\pi t_{\rm esc})^{-1}$, so if $t_{lc}=t_{\rm esc}$, the break will still
be at a frequency $2\pi$ lower than the one related to the light
crossing timescale.  Furthermore, the escape timescale could in
principle be longer than the light crossing timescale, since magnetic
fields in the blob would curve the electron's path and decrease the
time it takes to escape.  We note that in Figure \ref{PSD_flux_fig},
the break in the 0.1 GeV PSD is indeed associated with the escape
timescale, $f=(2\pi t_{\rm esc})^{-1}$, showing that this is at least
plausible.  If the break in the PSD of \object{3C 454.3}
\citep{nakagawa13} is due to electron escape, then the escape
timescale in the observer's frame will be $t_{\rm esc}=7.9$\ days$/ (2\pi)
= 30$\ hours, and in the comoving frame, 
\begin{eqnarray}
t_{\rm esc}^\prime = 20\ \ddays \left(\frac{\delta_D}{30}\right)\ . 
\end{eqnarray}

How could one distinguish between these interpretations?  One
possibility would be to observe the PSDs at more than one waveband.
If the break is due to the light-crossing timescale, the break
frequency should be present independent of the waveband.  The escape
timescale break could also be independent of frequency if the escape
timescale is energy independent, as it is in our model.  The cooling
timescale should be energy-dependent, and thus the break frequency
will be different in different wavebands.  For 3C 454.3, the
light-crossing timescale interpretation is disfavored since smaller
timescale fluctuations are present in its light curve
\citep[e.g.,][]{ackermann10_3c454.3,abdo11_3c454.3}.

\subsection{Optical PSDs of Blazars}

\citet{chatterjee12} compute $R$ band PSDs for 6 blazars based on
about 200-250 days of continuous data.  Their PSDs have power-law
indices are significantly steeper than those from the same objects'
$\gamma$-ray PSDs from \citet{nakagawa13}.  The exception is PKS
1510$-$089, for which \citet{chatterjee12} compute
$b=0.6^{+0.2}_{-0.5}$, significantly flatter than the $\gamma$-ray
PSD.  Our theory predicts that synchrotron and EC emission should have
the same PSD slopes if produced by the same electron energies, and all
but one of their sources are FSRQs, which are expected to emit
$\g$-rays by EC.  One possible reason for the discrepancy could be the
contamination in the optical by the accretion disk.  Another
possibility is that the time intervals used by \citet{chatterjee12}
are significantly shorter than the ones used by \citet{nakagawa13}.
As \citet{chatterjee12} point out, the large number of bright flares
in their time interval for PKS 1510$-$089 could be the cause of its
especially flat $R$ band PSD power-law index.  It could also be that
the different analysis methods used by \citet{chatterjee12} and
\citet{nakagawa13} could lead to different results.  Finally, it could
be that one of the assumptions of our theory is just not correct.

The {\em Kepler} mission, with its excellent relative photometry and
short timescale sampling is well-suited for measuring high-frequency
PSDs.  \citet{wehrle13} reported the {\em Kepler} PSDs of several
radio-loud AGN, and found no departure from a single power-law up to
$\sim 10^{-5}$\ Hz, above which white noise dominates.
\citet{edelson13} explored the {\em Kepler} PSD of the BL Lac object
\object{W2R1926+42} and found a ``bending'' power-law provided a good
fit to its PSD, with ``bend frequency'' corresponding to $\approx$ 4
hours.  The source \object{W2R1926+42} has a synchrotron peak at
$10^{14.5}$\ Hz according to \citet{edelson13}, making it an
intermediate synchrotron peaked (ISP) object by the classification of
\citet{abdo10_sed}.  However, its optical SED appears to be dominated
by accretion disk emission, implying its synchrotron peak is probably
at $\la 10^{13.5}$\ Hz, which would make it an LSP.  The {\em Kepler}
light curve could have a contribution from both the thermal accretion
disk emission and the nonthermal jet emission, making interpretation
of its PSD difficult.  If the optical band is dominated by synchrotron
emission, its status as an LSP implies that the electrons that produce
its optical emission are in the regime $\gp \gg (\nu
t_{\rm esc}^\prime)^{-1}$, meaning the ``bend frequency'' could be
associated with the light-crossing timescale or the cooling timescale.
If it is associated with the light-crossing timescale, the size of the
emitting region is
\begin{eqnarray}
R^\prime \approx 5.5\times10^{15}\left( \frac{\dD}{30}\right)\ \cm\ .
\end{eqnarray}

If it is associated with the cooling timescale, the cooling is 
dominated by EC, and $\dD=\G$, then 
\begin{flalign}
u_0 & \approx 3.2\times10^{-5}\ \left( \frac{\G}{30} \right)^{-5/2}
\left( \frac{B}{1\ \Gauss}\right)^{1/2}
\nonumber \\ & \times
\left( \frac{\lambda_{\rm obs}}{5000\ {\rm \AA}} \right)^{-1/2}\ 
\erg\ \cm^{-3}\ 
\end{flalign}
where $\lambda_{\rm obs}$ is the observed wavelength.  If the cooling is 
dominated by synchrotron, then the cooling timescale can be used to 
estimate the magnetic field, 
\begin{eqnarray}
B \approx 0.81\ \left( \frac{\dD}{30} \right)^{-1/3} 
\left( \frac{\lambda_{\rm obs}}{5000\ {\rm \AA}}\right)^{-1/3} 
\ \Gauss\ .
\end{eqnarray}

\subsection{The X-ray PSD of 3C~273}

An X-ray PSD of 3C~273 based on data combined from {\em RXTE}, {\em
EXOSAT} and other instruments was reported by \citet{mchardy08}.
Similar to the $\g$-ray PSD of 3C~454.3, the PSD shows two power-laws
with $b_1\approx 1.1$, $b_2\approx2.9$, with $f_{brk}\approx 1.0\times
10^{-6}\ \Hz$.  The break is close to 2, and the break frequency
corresponds to a timescale of $1.0\times 10^6\ \s \approx 280$\ hours
$\approx 12$\ days.  The interpretation for this break is more
difficult than the $\g$-ray PSD for 3C 454.3, since it is not clear
whether the X-rays are produced by SSC, EC, or even by a hot corona at
the base of the jet.  If the X-ray emission is dominated by EC, the
most likely interpretation of the break is with the escape timescale,
since the electrons generating the EC emission will almost certainly
have Lorentz factors $\g \ll (\nu t_{\rm esc})^{-1}$.  In this case the
escape timescale in the observer's frame is $t_{\rm esc} = 12$\
days$/(2\pi) = 46$\ hours.  

\subsection{Quasi-Periodic Oscillations}

A number of Quasi-periodic oscillations (QPOs) have been reported in
the X-ray and optical PSDs of blazars.  These QPOs could be associated
with the maxima at half-integer values of either the light crossing
timescale ($t_{lc}$) or the cooling timescale ($t_{\rm cool}$).  See
for example Figure \ref{PSD_flux_fig}, where maxima in the 0.648
$\mu$m or 1.0 GeV PSDs could be confused with QPOs in noisy PSDs.  If
this interpretation is correct, one could distinguish between these
possibilities ($t_{lc}$ or $t_{\rm cool}$) by observing the PSDs at
more than one wavelength.  If the QPO appears in at the same frequency
independent of wavelength, it would argue for a $t_{lc}$
interpretation.  If QPOs are found at different frequencies at
different wavelengths, it argues for the $t_{\rm cool}$
interpretation.  Assessing the significance of QPOs in
red noise PSDs can be subtle \citep[e.g.,][]{vaughan05,vaughan06}.

As an example, we look at the claimed QPO in reported in {\em
XMM-Newton} observations of \object{PKS 2155$-$304}
\citep{lachowicz09}.  Visually inspecting their PSD from data taken
between 0.3 and 10 keV, one sees significant maxima at $\sim
3.5\times10^{-5}$\ Hz and $\sim 7.0\times10^{-5}$\ Hz.
\citet{lachowicz09} also examined the PSDs in the energy bands 0.3--2
keV and 2--10 keV, and found the maxima were significant in the soft
band but not in the hard band, although they were still found in the
hard band.  \citet{gaur10} and \citet{gonzalez12} find this QPO
in only one of many {\em XMM-Newton} observations of this source.  If
the maxima are significant and found in both energy bands, this
argues for a light crossing time interpretation, with timescale
$t_{lc}=3.8\times10^{4}\ \s$ and
\begin{eqnarray}
R^\prime \approx \frac{c t_{lc} \dD}{2(1+z)} = 3.1\times10^{16}\left( \frac{\dD}{60}\right)\ \cm\ .
\end{eqnarray}
This is larger than the size from the HESS observation (Section
\ref{VHEPSD}).  The {\em XMM-Newton} observations were taken on 2006
May 1, and the HESS observations on 2006 July 28, which could account
for the discrepancy.  The emitting region size could have been
different at different times.  A number of claimed significant
detections of QPOs from the literature are listed in Table
\ref{QPO_table}.

\begin{deluxetable*}{lccc}
\tabletypesize{\scriptsize}
\tablecaption{
Claimed QPOs from blazars reported in the literature.
}
\tablewidth{0pt}
\tablehead{
\colhead{Authors} &
\colhead{Object} &
\colhead{Bandpass} &
\colhead{QPO frequency [Hz]} 
}
\startdata
\citet{espaillat08}\footnotemark[1] & \object{3C 273} & 0.75--10 keV & $3.0\times10^{-4}$ \\
\citet{lachowicz09} & \object{PKS 2155$-$304} & 0.2--10 keV & $6\times10^{-5}$ \\
\citet{gupta09} & \object{S5 0716+714} & $V$ and $R$ band & various \\
\citet{rani10} & \object{S5 0716+714} & $R$ band & $1.1\times10^{-3}$ 
\enddata
\label{QPO_table}
\footnotetext[1]{This claimed QPO is disputed by \citet{mohan11} and \citet{gonzalez12}.}
\end{deluxetable*}

\section{Discussion}
\label{discussion}

We have presented a new theoretical formalism for modeling the
variability of blazars, based on an analytical solution to the
underlying electron continuity equation governing the distribution of
radiating electrons in a homogeneous blob moving out in the jet. The
analysis was carried out in the Fourier domain, and the results are
therefore directly comparable with observational Fourier data products
such as the PSDs and time/phase lags. This formalism assumes emission
from a jet closely aligned with the line of sight, so that the
emission produced in the comoving frame is Doppler shifted into the
observer's frame.  Internal shocks in the jet randomly accelerate
electrons to high energies, which are then injected into an emitting
region at random intervals with that variability characterized by a
power-law in Fourier space.  The observable radiation produced by
the electrons is affected by cooling, electron escape, and the
light-travel time across the blob. The model makes specific
predictions regarding the the PSDs and Fourier-dependent time lag
components resulting from synchrotron, EC, and SSC emission, and it
successfully reproduces the characteristics of the colored noise seen
in nearly all blazars.

The study presented here is a first attempt at examining blazar 
variability with this formalism.  As such, it makes a number of 
simplifying assumptions.  
\begin{enumerate}
\item It assumes the only thing which varies with time in a blazar is
the rate at which electrons are injected into the emitting region.
All other parameters---the magnetic field strength ($B$), the size of
the emitting region ($R^\prime$), the electron injection power-law
index ($q$), the jet's angle to the line of sight ($\theta$), and so
on---are assumed not to vary.  Although this is likely an
over-simplification, we note that the PSDs of PKS 0537$-$441 can be
explained by only varying the electron distribution of the source
\citep{dammando13_0537}, so that in some cases this may be justified.
\item In computing emission, we have used simplified $\delta$-function
expressions for the synchrotron and Compton emission, and assumed
Thomson scattering.  Using more accurate expressions, in particular,
the full Compton cross-section for the energy losses could lead to
interesting effects \citep{dermer02_KN,moderski05,sikora09,dotson12}.  
\item The calculations neglect SSC cooling, which is quite difficult
to model analytically \citep{schlickeiser09,zacharias10,zacharias12,
zacharias12_EC}.  This would likely not be important for FSRQs, where
the EC component probably dominates the cooling, but could be
important for BL Lac objects that do not have a strong external
radiation field.
\item We have neglected the details of the acceleration
mechanism.  Such a mechanism may produce interesting features in PSDs
that could be observable.
\item Although we take light travel-time effects into account, we
assume all parts of the blob vary simultaneously, which is obviously
not the case.
\end{enumerate}
In the future, we will perform more detailed analyses which explore
these more complicated cases.

What are the best wavebands to observe blazars and compare with the
theory outlined in this work?  Observations at (electromagnetic)
frequencies lower than $\sim 10^{12}\ \Hz$ would not be useful, since
here the emission is likely dominated by the superposition of many
self-absorbed jet components \citep{konigl81}.  For low-synchrotron
peaked (LSP) blazars \citep[including almost all FSRQs;][]{finke13},
The optical and GeV $\g$-rays would have features at high Fourier
frequencies due the rapid energy losses of electrons that produce this
radiation, and thus this emission could be quite interesting to
observe.  Observing such short timescales with the {\em Fermi}-LAT
could be difficult, since for this instrument one must usually
integrate over fairly long timescales ($\ga$ a few hours for all but
the brightest sources) to get a significant detection.  Bright flares
and adaptive light curve binning may be helpful in producing accurate
LAT PSDs at high Fourier frequencies \citep{lott12}.  Significant
detections can be made in the optical with shorter integration times
($\sim$ a few minutes or even fractions of a minute), leading to
better PSDs at high frequencies \citep[e.g.,][]{rani10} although in
FSRQs this could be contaminated by thermal emission from an accretion
disk.

Observing LSPs at wavebands (e.g., infrared) where emission is
dominated by less energetic electrons could probe the escape
timescale.  PSDs produced from simultaneous light curves at multiple
frequencies would be extremely helpful for verifying the predictions
of this paper.  At low Fourier frequencies the PSDs should have
essentially identical power-law shapes at all wavebands, and at higher
Fourier frequencies features associated with the light-crossing
timescale could be identified, and should be the same in all
wavebands.  This could serve as a strong test for the theory presented
in this paper.  We also predict that all breaks in observed
synchrotron and EC PSD power laws should be by 2, i.e., from $\propto
f^{-a}$ to $\propto f^{-(a+2)}$, and more gradual breaks could be
observed in SSC PSDs.  \citet{kataoka01} observed smaller breaks in
the X-ray PSDs of Mrk 421, Mrk 501, and PKS 2155$-$304 based on {\em
ASCA} and {\em Rossi X-ray Timing Explorer} observations.  However,
note that they did not obtain acceptable fits to their PSDs with
broken power-laws with all parameters left free to vary.  Also, it may
be that including SSC cooling could modify the the PSD so that a
breaks other than 2 are possible, although that is beyond the scope of
this paper.

For high synchrotron-peaked blazars (HSPs), almost all of which are BL
Lac objects, the $\g$-ray emission is expected to be from SSC.  We do
not predict any features in SSC PSDs from cooling or escape, although
features from the light crossing time are still expected.  These
predictions could be tested with {\em Fermi}-LAT and VHE $\g$-ray
instruments such as MAGIC, HESS, VERITAS, and the upcoming CTA.  For
synchrotron emission, however, features from cooling and escape should
be present.  An X-ray telescope could be used to probe emission from
the highest energy electrons potentially seeing cooling features.  The
proposed {\em Large Observatory for X-ray Timing (LOFT)} spacecraft
could be very useful for this \citep{donnarumma13}.  {\em LOFT} will
provide excellent timing coverage ($\sim$\ ms) with a relatively large
effective area.  PSDs produced from optical light curves of HSPs could
be used to probe the escape timescale.  As with LSP blazars, PSDs
produced by simultaneous light curves from multiple wavebands would be
extremely helpful for the study of HSP blazars.

\acknowledgements 

We are grateful to the anonymous referee for insightful
suggestions that helped improve the discussion and presentation, and
to C.\ Dermer for useful discussions.  J.D.F.\ was supported by the
Office of Naval Research.

\appendix

\section{Solution to the Fourier Transform of the Continuity Equation}
\label{FTCEsoln}

Here we solve the differential equation
\begin{eqnarray}
\label{fourier_cont_eqn2}
-2\pi i f \tilN_e(\g,f) + \frac{\partial}{\partial\g} [\dotg(\g) \tilN_e(\g,f)] + 
\frac{\tilN_e(\g,f)}{t_{\rm esc}(\g)} = 
\tilQ(\g; f)\ 
\end{eqnarray}
for $\dotg\le 0$.  This equation can be rearranged, 
\begin{eqnarray}
\frac{\partial}{\partial\g} [\dotg(\g) \tilN_e(\g,f)] + 
\left[ \frac{1}{t_{\rm esc}} - i\omega \right] \tilN_e(\g,f) = 
\tilQ(\g; f)\ 
\end{eqnarray}
recalling that $\omega=2\pi f$.  One can multiply both sides by
\begin{eqnarray}
\exp\left[ \int^\infty_\g \frac{d\gp}{ |\dotg(\gp)| }
\left( \frac{1}{t_{\rm esc}} - i\omega \right) \right]
\end{eqnarray}
and then rearrange the left side so that
\begin{eqnarray}
\frac{d}{d\g} 
\left\{ 
\dotg(\g) \tilN_e(\g,f) 
\exp\left[ \int^{\infty}_\g \frac{d\gp}{|\dot{\g}(\gp)|}
\left( \frac{1}{t_{\rm esc}} - i\omega \right) \right]  
\right\}
\nonumber \\ 
= \tilQ(\g; f) 
\exp\left[ \int^\infty_\g \frac{d\gp}{|\dotg(\gp)|}
\left( \frac{1}{t_{\rm esc}} - i\omega \right) \right]\  .
\end{eqnarray}
Integrating both sides gives
\begin{eqnarray}
|\dotg(\g)| \tilN_e(\g,f) 
\exp\left[ \int^\infty_\g \frac{d\gp}{|\dotg(\gp)|}
\left( \frac{1}{t_{\rm esc}} - i\omega \right) \right]  
\nonumber \\ 
= \int_\g^\infty d\gp\ \tilQ(\gp; f) 
\exp\left[ \int^{\infty}_{\gp} \frac{d\gamma^{\prime\prime}}{|\dotg(\g^{\prime\prime})|}
\left( \frac{1}{t_{\rm esc}} - i\omega \right) \right]\  .
\end{eqnarray}
Solving this for $\tilN(\g,f)$ results in
\begin{eqnarray}
\tilN_e(\g,f) = \frac{1}{|\dotg(\g)|}\int_\g^{\infty} d\gp\ \tilQ(\gp;f)\ 
\exp\left[ -\int_{\g}^{\gp} \frac{d\g^{\prime\prime}}{|\dotg(\g^{\prime\prime})|}
\left(\frac{1}{t_{\rm esc}(\g^{\prime\prime})} - i\omega\right)\right]\ .
\end{eqnarray}

\section{Normalization of Electron Injection Function}
\label{normalization}

The time average of the total power injected in electrons is
\begin{eqnarray}
\langle L_{\rm inj}\rangle = \frac{m_e c^2}{\Delta t} \int^\infty_{1} d\g\ \g\ 
\int_{0}^{\Delta t} dt\ Q(\g,t)
\end{eqnarray}
where $\Delta t$ is the length of the time interval over which the electrons 
are injected.  
Using Equation (\ref{IFT_define}), 
\begin{eqnarray}
\langle L_{\rm inj}\rangle = \frac{m_e c^2}{\Delta t} \int^\infty_{1} d\g\ \g\  
\int^\infty_{-\infty} df\  \tilQ(\g,f) \int_{0}^{\Delta t}  dt\ \exp[-2\pi i f t]\ .
\end{eqnarray}
Substituting Equation (\ref{tilQ}), 
\begin{eqnarray}
\tilQ(\g;f) = Q_0 (f/f_0)^{-a/2}\g^{-q}H(\g;\g_1,\g_2) H(f; f_1, f_2) \ ,
\end{eqnarray}
and integrating over $\g$, one gets
\begin{eqnarray}
\langle L_{\rm inj}\rangle = \frac{m_e c^2}{\Delta t} Q_0\  
G(q,\g_1,\g_2)\ 
\int^{f_2}_{f_1} df\  (f/f_0)^{-a/2} \int_{0}^{\Delta t}  dt\ \exp[-2\pi i f t]\ 
\end{eqnarray}
where
\begin{equation}
G(q, \g_1, \g_2) = 
\left\{ \begin{array}{ll}
(\g_1^{2-q}-\g_2^{2-q})/(q-2) & \mathrm{for}\ q \ne 2 \\
 \ln(\g_2/\g_1)  & \mathrm{for}\ q = 2
\end{array}
\right. \ .
\end{equation}
Performing the integral over time gives
\begin{eqnarray}
\langle L_{\rm inj}\rangle = \frac{m_e c^2}{-2\pi i \Delta t} Q_0\  
G(q,\g_1,\g_2)\ 
\int^{f_2}_{f_1} \frac{df}{f}\  (f/f_0)^{-a/2} 
\left[ \exp(-2\pi i f \Delta t) - 1\right]\ .
\end{eqnarray}
This can be rewritten as
\begin{eqnarray}
\label{Linj}
\langle L_{\rm inj}\rangle = \frac{m_e c^2}{-2\pi i \Delta t} Q_0\  
G(q,\g_1,\g_2)\ [I_r(a,f_1,f_2) - i I_i(a,f_1,f_2) - I_0(a,f_1,f_2)]
\end{eqnarray}
where
\begin{eqnarray}
I_r(a,f_1,f_2) = \int^{f_2}_{f_1} \frac{df}{f}\  (f/f_0)^{-a/2} \cos(2\pi f \Delta t)\ ,
\end{eqnarray}
\begin{eqnarray}
I_i(a,f_1,f_2) = \int^{f_2}_{f_1} \frac{df}{f}\  (f/f_0)^{-a/2} \sin(2\pi f \Delta t)\ ,
\end{eqnarray}
and
\begin{equation}
I_0(a,f_1,f_2) = \int^{f_2}_{f_1} \frac{df}{f}\  (f/f_0)^{-a/2}  = 
\left\{ \begin{array}{ll}
2/a[(f_1/f_0)^{-a/2} - (f_2/f_0)^{-a/2} ] & a\ne0 \\
\ln(f_2/f_1) & a = 0
\end{array}
\right. \ .
\end{equation}
Multiplying Equation (\ref{Linj}) by its complex conjugate gives
\begin{eqnarray}
|\langle L_{\rm inj}\rangle|^2 = \left[ \frac{m_e c^2 Q_0 G }{2\pi \Delta t} \right]^2
[I_r^2 + I_i^2 - 2 I_r I_0 + I_0^2]\ ,
\end{eqnarray}
or, solving for $Q_0$,
\begin{eqnarray}
Q_0 = \frac{2\pi \Delta t \langle L_{\rm inj}\rangle}{m_e c^2 G
\sqrt{ I_r^2 + I_i^2 - 2 I_r I_0 + I_0^2} }\ .
\end{eqnarray}

\section{Light Travel Time}
\label{lighttravelappendix}

\begin{figure}
\vspace{2.2mm} 
\epsscale{0.4} 
\plotone{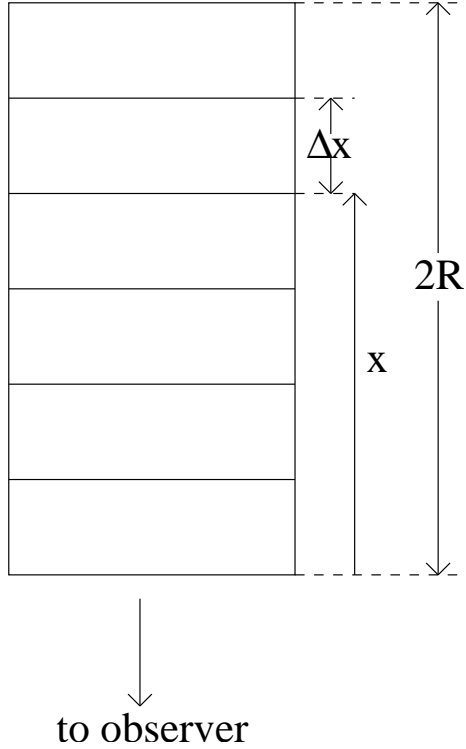}
\caption{ Sketch of geometry of emitting blob for the purpose of
computing light travel time effects.  This sketch is in the frame co-moving 
with the blob.  The blob of length $2R$ is divided into $N$ pieces each with 
length $\Delta x$.}
\label{varfig}
\vspace{2.2mm}
\end{figure}

Let us take a cylindrical blob, as shown in cross section in Figure \ref{varfig}.  The
blob has length $2R$, and everywhere within the blob radiation is
emitted simultaneously as a function of time $t$ as $g(t)$.  The
entire blob has length $R$ and is divided into $N$ individual
segments, each with length $\Delta x=2R/N$.  The radiation emitted by
each individual segment is a corresponding fraction of the whole,
$g(t)/N = g(t) \Delta x/2R$.  The radiation an observer co-moving with
the blob sees at any given time $t_{\rm obs}$ will be a sum over the
individual segments at that time,
\begin{eqnarray}
h(t_{\rm obs}) = g(t_{\rm obs}-x_1/c)\ \Delta x/(2R) + g(t_{\rm obs}-x_2/c)\ \Delta x/(2R) + 
\\ \nonumber
\ldots + g(t_{\rm obs}-2R/c)\ \Delta x/(2R)
\end{eqnarray}
or
\begin{eqnarray}
h(t_{\rm obs}) = \frac{1}{2R}\ \sum_{j=1}^N\  g(t_{\rm obs}-x_j/c)\ \Delta x\ .
\end{eqnarray}
As $N \rightarrow \infty$, 
\begin{eqnarray}
h(t_{\rm obs}) \rightarrow \frac{1}{2R} \int_0^{2R}\ dx\  g(t_{\rm obs}-x/c)\ .
\end{eqnarray}
Since $t=x/c$, 
\begin{eqnarray}
h(t_{\rm obs}) = \frac{c}{2R} \int_0^{2R/c}\ dt\  g(t_{\rm obs}-t)\ .
\end{eqnarray}

Let us now move to a frame where the blob is moving relative to the
observer so that she or he sees a time $t_{\rm obs}=t_{\rm obs}\p(1+z)/\dD$,
where now all lengths and times in the co-moving frame will be primed.
Then
\begin{eqnarray}
h(t_{\rm obs}) = \frac{c}{2R\p} \int_0^{2R\p/c}\ dt\p\  g(t\p_{\rm obs}-t\p)\ 
\\ \nonumber
= \frac{c}{2R\p} \int_0^{2R\p/c}\ dt\p\  g\left(\frac{t_{\rm obs}\dD}{1+z}-t\p\right)\ .
\end{eqnarray}

This is similar to the ``time slices'' of \citet{chiaberge99}.  We use
a cylindrical normal blob geometry for simplicity here, although a
spherical one would be more consistent with the SSC calculation.
Since the actual geometry of an emitting blob is not known,
differences should not be too great.  A similar derivation for a
spherical geometry is given by \citet{zacharias13}.

\section{Fourier Transform Including Light Travel Time}
\label{lighttravelfourierappendix}

In this appendix we derive Equation (\ref{tilF5}) from Equation
(\ref{Fsy1}).  From the definition of the inverse Fourier transform, 
Equation (\ref{IFT_define}), 
\begin{eqnarray}
N_e(\g; t) = \int^{\infty}_{-\infty} df \tilN_e(\g,f) \exp(-2\pi i f t)\ .
\end{eqnarray}
Putting this in Equation (\ref{Fsy1}) and rearranging gives
\begin{eqnarray}
\label{Fsy2}
F^{sy}_\e(t) = \frac{K(1+z)}{t_{lc}\dD} \int^{\infty}_{-\infty} df^\prime \tilN_e(\gp,f^\prime) 
\exp\left[-\frac{2\pi i \fpr \delta_D t}{1+z}\right] 
\int^{2R^\prime/c}_0 d\tp \exp(2\pi i \fpr \tp)\ ,
\end{eqnarray}
recalling that $t_{lc}=2R^\prime (1+z)/(c \dD)$.  Performing the integral
over $\tp$ gives
\begin{eqnarray}
\label{Fsy3}
F^{sy}_\e(t) = \frac{K(1+z)}{2\pi i t_{lc} \dD} \int^{\infty}_{-\infty} \frac{df^\prime}{\fpr}
\tilN_e(\gp,f^\prime) \exp\left[ \frac{-2\pi i \fpr \dD t}{1+z}\right] 
\left\{ \exp\left[ \frac{4 \pi i \fpr R^\prime}{c}\right] - 1 \right\}\ .
\end{eqnarray}
The Fourier transform of the synchrotron flux light curve is defined as
(see Equation [\ref{FT_define}])
\begin{eqnarray}
\label{tilF1}
\tilF^{sy}_\e(f) = \int^{\infty}_{-\infty} dt\ F_\e(t) \exp(2\pi i f t)\ .
\end{eqnarray}
Substituting Equation (\ref{Fsy3}) for $F_\e(t)$ in this equation gives
\begin{eqnarray}
\label{tilF2}
\tilF^{sy}_\e(f) = \frac{K(1+z)}{2\pi i t_{lc} \dD} \int^{\infty}_{-\infty} \frac{df^\prime}{\fpr}
\tilN_e(\gp, \fpr) \left\{ \exp\left[ \frac{4 \pi i \fpr R^\prime}{c}\right] - 1 \right\} 
\\ \nonumber \times 
\int^{\infty}_{-\infty} dt 
\exp\left[2\pi it \left(f - \frac{\fpr\dD}{1+z}\right)\right]\ .
\end{eqnarray}
The integral over $t$ has the form of a Dirac $\delta$-function
(Equation [\ref{dirac}]), so
\begin{eqnarray}
\label{tilF3}
\tilF^{sy}_\e(f) = \frac{K(1+z)}{2\pi i t_{lc}\dD} \int^{\infty}_{-\infty} \frac{df^\prime}{\fpr}
\tilN_e(\gp, \fpr) \left\{ \exp\left[ \frac{4 \pi i \fpr R^\prime}{c}\right] - 1 \right\}
\delta\left( f - \frac{\fpr\dD}{1+z}\right)\ .
\end{eqnarray}
Using the well-known property for $\delta$ functions,
\begin{eqnarray}
\delta\left( f - \frac{\fpr\dD}{1+z}\right) = 
\frac{1+z}{\dD}\delta\left(\fpr - \frac{(1+z)f}{\dD}\right)\ ,
\end{eqnarray}
one can perform the integral over $\fpr$ in Equation (\ref{tilF3}) to get
\begin{eqnarray}
\label{tilF4}
\tilF^{sy}_\e(f) = \frac{K(1+z)}{2\pi if t_{lc}\dD} \tilN_e\left(\gp, \frac{(1+z)f}{\dD}\right)\ 
\left\{ \exp\left[ \frac{4 \pi i f(1+z) R^\prime}{c\dD}\right] - 1 \right\}\ .
\end{eqnarray}
This is Equation (\ref{tilF5}).


\bibliographystyle{apj}
\bibliography{variability_ref,EBL_ref,references,mypapers_ref,blazar_ref,sequence_ref,SSC_ref,LAT_ref,3c454.3_ref}

\begin{thebibliography}{67}
\expandafter\ifx\csname natexlab\endcsname\relax\def\natexlab#1{#1}\fi

\bibitem[{{Abdo} {et~al.}(2010{\natexlab{a}}){Abdo}, {Ackermann}, {Ajello},
  {Allafort}, {Antolini}, {Atwood}, {Axelsson}, {Baldini}, {Ballet},
  {Barbiellini}, \& et~al.}]{abdo10_1fgl}
{Abdo}, A.~A., {et~al.} 2010{\natexlab{a}}, \apjs, 188, 405

\bibitem[{{Abdo} {et~al.}(2010{\natexlab{b}}){Abdo}, {Ackermann}, {Ajello},
  {Antolini}, {Baldini}, {Ballet}, {Barbiellini}, {Bastieri}, {Bechtol},
  {Bellazzini}, {Berenji}, {Blandford}, {Bloom}, {Bonamente}, {Borgland},
  {Bouvier}, {Bregeon}, {Brez}, {Brigida}, {Bruel}, {Buehler}, {Burnett},
  {Buson}, {Caliandro}, {Cameron}, {Caraveo}, {Carrigan}, {Casandjian},
  {Cavazzuti}, {Cecchi}, {{\c C}elik}, {Chekhtman}, {Cheung}, {Chiang},
  {Ciprini}, {Claus}, {Cohen-Tanugi}, {Cominsky}, {Conrad}, {Costamante},
  {Cutini}, {Dermer}, {de Angelis}, {de Palma}, {Silva}, {Drell}, {Dubois},
  {Dumora}, {Farnier}, {Favuzzi}, {Fegan}, {Focke}, {Fortin}, {Frailis},
  {Fukazawa}, {Funk}, {Fusco}, {Gargano}, {Gasparrini}, {Gehrels}, {Germani},
  {Giebels}, {Giglietto}, {Giommi}, {Giordano}, {Glanzman}, {Godfrey},
  {Grenier}, {Grondin}, {Grove}, {Guiriec}, {Hadasch}, {Hayashida}, {Hays},
  {Healey}, {Horan}, {Hughes}, {Itoh}, {J{\'o}hannesson}, {Johnson}, {Johnson},
  {Kamae}, {Katagiri}, {Kataoka}, {Kawai}, {Kn{\"o}dlseder}, {Kuss}, {Lande},
  {Larsson}, {Latronico}, {Lemoine-Goumard}, {Longo}, {Loparco}, {Lott},
  {Lovellette}, {Lubrano}, {Madejski}, {Makeev}, {Massaro}, {Mazziotta},
  {McEnery}, {Michelson}, {Mitthumsiri}, {Mizuno}, {Moiseev}, {Monte},
  {Monzani}, {Morselli}, {Moskalenko}, {Mueller}, {Murgia}, {Nolan}, {Norris},
  {Nuss}, {Ohno}, {Ohsugi}, {Omodei}, {Orlando}, {Ormes}, {Ozaki}, {Panetta},
  {Parent}, {Pelassa}, {Pepe}, {Pesce-Rollins}, {Piron}, {Porter}, {Rain{\`o}},
  {Rando}, {Razzano}, {Reimer}, {Reimer}, {Ritz}, {Rodriguez}, {Romani},
  {Roth}, {Ryde}, {Sadrozinski}, {Sander}, {Scargle}, {Sgr{\`o}}, {Shaw},
  {Smith}, {Spandre}, {Spinelli}, {Starck}, {Strickman}, {Suson}, {Takahashi},
  {Takahashi}, {Tanaka}, {Thayer}, {Thayer}, {Thompson}, {Tibaldo}, {Torres},
  {Tosti}, {Tramacere}, {Uchiyama}, {Usher}, {Vasileiou}, {Vilchez}, {Vitale},
  {Waite}, {Wallace}, {Wang}, {Winer}, {Wood}, {Yang}, {Ylinen}, \&
  {Ziegler}}]{abdo10_var}
---. 2010{\natexlab{b}}, \apj, 722, 520

\bibitem[{{Abdo} {et~al.}(2010{\natexlab{c}})}]{abdo10_sed}
---. 2010{\natexlab{c}}, \apj, 716, 30

\bibitem[{{Abdo} {et~al.}(2011{\natexlab{a}}){Abdo}, {Ackermann}, {Ajello},
  {Allafort}, {Baldini}, {Ballet}, {Barbiellini}, {Bastieri}, {Bellazzini},
  {Berenji}, {Blandford}, {Bloom}, {Bonamente}, {Borgland}, {Bouvier},
  {Bregeon}, {Brigida}, {Bruel}, {Buehler}, {Buson}, {Caliandro}, {Cameron},
  {Caraveo}, {Casandjian}, {Cavazzuti}, {Cecchi}, {Charles}, {Chekhtman},
  {Cheung}, {Chiang}, {Ciprini}, {Claus}, {Conrad}, {Cutini}, {D'Ammando}, {de
  Angelis}, {de Palma}, {Dermer}, {Digel}, {Silva}, {Drell}, {Dubois},
  {Dumora}, {Escande}, {Favuzzi}, {Fegan}, {Ferrara}, {Fortin}, {Fukazawa},
  {Fusco}, {Gargano}, {Gasparrini}, {Gehrels}, {Germani}, {Giglietto},
  {Giommi}, {Giordano}, {Giroletti}, {Glanzman}, {Godfrey}, {Grenier}, {Grove},
  {Guiriec}, {Hadasch}, {Hayashida}, {Hays}, {Horan}, {Itoh},
  {J{\'o}hannesson}, {Johnson}, {Kamae}, {Katagiri}, {Kataoka},
  {Kn{\"o}dlseder}, {Kuss}, {Lande}, {Larsson}, {Latronico}, {Lee}, {Longo},
  {Loparco}, {Lott}, {Lovellette}, {Lubrano}, {Madejski}, {Makeev},
  {Mazziotta}, {McConville}, {McEnery}, {Michelson}, {Mitthumsiri}, {Mizuno},
  {Moiseev}, {Monte}, {Monzani}, {Morselli}, {Moskalenko}, {Murgia},
  {Naumann-Godo}, {Nishino}, {Nolan}, {Norris}, {Nuss}, {Ohsugi}, {Okumura},
  {Orlando}, {Ormes}, {Paneque}, {Pelassa}, {Pesce-Rollins}, {Pierbattista},
  {Piron}, {Porter}, {Rain{\`o}}, {Rando}, {Razzaque}, {Reimer}, {Reimer},
  {Ritz}, {Roth}, {Sadrozinski}, {Sanchez}, {Scargle}, {Schalk}, {Sgr{\`o}},
  {Siskind}, {Smith}, {Spandre}, {Spinelli}, {Strickman}, {Takahashi},
  {Takahashi}, {Tanaka}, {Tanaka}, {Thayer}, {Thayer}, {Thompson}, {Tibaldo},
  {Torres}, {Tosti}, {Tramacere}, {Troja}, {Vandenbroucke}, {Vasileiou},
  {Vianello}, {Vilchez}, {Vitale}, {Waite}, {Wang}, {Winer}, {Wood}, {Yang}, \&
  {Ziegler}}]{abdo11_3c454.3}
---. 2011{\natexlab{a}}, \apjl, 733, L26

\bibitem[{{Abdo} {et~al.}(2011{\natexlab{b}}){Abdo}, {Ackermann}, {Ajello},
  {Baldini}, {Ballet}, {Barbiellini}, {Bastieri}, {Bechtol}, {Bellazzini},
  {Berenji}, \& et~al.}]{abdo11_mrk421}
---. 2011{\natexlab{b}}, \apj, 736, 131

\bibitem[{{Abdo} {et~al.}(2011{\natexlab{c}}){Abdo}, {Ackermann}, {Ajello},
  {Allafort}, {Baldini}, {Ballet}, {Barbiellini}, {Baring}, {Bastieri},
  {Bechtol}, \& et~al.}]{abdo11_mrk501}
---. 2011{\natexlab{c}}, \apj, 727, 129

\bibitem[{{Ackermann} {et~al.}(2010){Ackermann}, {Ajello}, {Baldini}, {Ballet},
  {Barbiellini}, {Bastieri}, {Bechtol}, {Bellazzini}, {Berenji}, {Blandford},
  {Bonamente}, {Borgland}, {Bregeon}, {Brigida}, {Bruel}, {Buehler}, {Burnett},
  {Buson}, {Caliandro}, {Cameron}, {Caraveo}, {Carrigan}, {Casandjian},
  {Cavazzuti}, {Cecchi}, {{\c C}elik}, {Chekhtman}, {Cheung}, {Chiang},
  {Ciprini}, {Claus}, {Cohen-Tanugi}, {Corbel}, {Cutini}, {D'Ammando},
  {Dermer}, {de Angelis}, {de Palma}, {Digel}, {Silva}, {Drell}, {Dubois},
  {Dumora}, {Escande}, {Favuzzi}, {Fegan}, {Ferrara}, {Fuhrmann}, {Fukazawa},
  {Fusco}, {Gargano}, {Gasparrini}, {Gehrels}, {Germani}, {Giebels},
  {Giglietto}, {Giommi}, {Giordano}, {Giroletti}, {Glanzman}, {Godfrey},
  {Grenier}, {Grove}, {Guiriec}, {Hadasch}, {Hayashida}, {Hays},
  {J{\'o}hannesson}, {Johnson}, {Johnson}, {Kamae}, {Katagiri}, {Kataoka},
  {Kn{\"o}dlseder}, {Kuss}, {Lande}, {Larsson}, {Latronico}, {Lee}, {Llena
  Garde}, {Longo}, {Loparco}, {Lott}, {Lubrano}, {Madejski}, {Makeev},
  {Marchili}, {Mazziotta}, {McEnery}, {Mehault}, {Michelson}, {Mizuno},
  {Monte}, {Monzani}, {Morselli}, {Moskalenko}, {Murgia}, {Nakamori},
  {Nalewajko}, {Naumann-Godo}, {Nolan}, {Norris}, {Nuss}, {Ohsugi}, {Okumura},
  {Omodei}, {Orlando}, {Ormes}, {Pelassa}, {Pepe}, {Pesce-Rollins}, {Piron},
  {Porter}, {Rain{\`o}}, {Rando}, {Razzano}, {Reimer}, {Reimer}, {Reyes},
  {Ripken}, {Ritz}, {Roth}, {Sadrozinski}, {Sanchez}, {Sander}, {Scargle},
  {Sgr{\`o}}, {Sikora}, {Siskind}, {Spandre}, {Spinelli}, {Strickman}, {Suson},
  {Takahashi}, {Takahashi}, {Tanaka}, {Tanaka}, {Thayer}, {Thayer}, {Thompson},
  {Tibaldo}, {Torres}, {Tosti}, {Tramacere}, {Usher}, {Vandenbroucke},
  {Vilchez}, {Vitale}, {Waite}, {Wang}, {Wehrle}, {Winer}, {Yang}, {Ylinen}, \&
  {Ziegler}}]{ackermann10_3c454.3}
{Ackermann}, M., {et~al.} 2010, \apj, 721, 1383

\bibitem[{{Aharonian} {et~al.}(2007)}]{aharonian07_2155}
{Aharonian}, F., {et~al.} 2007, \apjl, 664, L71

\bibitem[{{Begelman} {et~al.}(2008){Begelman}, {Fabian}, \&
  {Rees}}]{begelman08}
{Begelman}, M.~C., {Fabian}, A.~C., \& {Rees}, M.~J. 2008, \mnras, 384, L19

\bibitem[{{B{\l}a{\.z}ejowski} {et~al.}(2000){B{\l}a{\.z}ejowski}, {Sikora},
  {Moderski}, \& {Madejski}}]{blazejowski00}
{B{\l}a{\.z}ejowski}, M., {Sikora}, M., {Moderski}, R., \& {Madejski}, G.~M.
  2000, \apj, 545, 107

\bibitem[{{Bloom} \& {Marscher}(1996)}]{bloom96}
{Bloom}, S.~D., \& {Marscher}, A.~P. 1996, \apj, 461, 657

\bibitem[{{B{\"o}ttcher} \& {Chiang}(2002)}]{boett02a}
{B{\"o}ttcher}, M., \& {Chiang}, J. 2002, \apj, 581, 127

\bibitem[{{B{\"o}ttcher} \& {Reimer}(2004)}]{boett04}
{B{\"o}ttcher}, M., \& {Reimer}, A. 2004, \apj, 609, 576

\bibitem[{{Chatterjee} {et~al.}(2012){Chatterjee}, {Bailyn}, {Bonning},
  {Buxton}, {Coppi}, {Fossati}, {Isler}, {Maraschi}, \& {Urry}}]{chatterjee12}
{Chatterjee}, R., {et~al.} 2012, \apj, 749, 191

\bibitem[{{Chen} {et~al.}(2012){Chen}, {Fossati}, {B{\"o}ttcher}, \&
  {Liang}}]{chen12}
{Chen}, X., {Fossati}, G., {B{\"o}ttcher}, M., \& {Liang}, E. 2012, \mnras,
  424, 789

\bibitem[{{Chen} {et~al.}(2011){Chen}, {Fossati}, {Liang}, \&
  {B{\"o}ttcher}}]{chen11}
{Chen}, X., {Fossati}, G., {Liang}, E.~P., \& {B{\"o}ttcher}, M. 2011, \mnras,
  416, 2368

\bibitem[{{Chiaberge} \& {Ghisellini}(1999)}]{chiaberge99}
{Chiaberge}, M., \& {Ghisellini}, G. 1999, \mnras, 306, 551

\bibitem[{{Cui}(2004)}]{cui04}
{Cui}, W. 2004, \apj, 605, 662

\bibitem[{{D'Ammando} {et~al.}(2013){D'Ammando}, {Antolini}, {Tosti}, {Finke},
  {Ciprini}, {Larsson}, {Ajello}, {Covino}, {Gasparrini}, {Gurwell}, {Hauser},
  {Romano}, {Schinzel}, {Wagner}, {Impiombato}, {Perri}, {Persic}, {Pian},
  {Polenta}, {Sbarufatti}, {Treves}, {Vercellone}, {Wehrle}, \&
  {Zook}}]{dammando13_0537}
{D'Ammando}, F., {et~al.} 2013, \mnras, 431, 2481

\bibitem[{{Dermer} \& {Atoyan}(2002)}]{dermer02_KN}
{Dermer}, C.~D., \& {Atoyan}, A.~M. 2002, \apjl, 568, L81

\bibitem[{{Dermer} \& {Menon}(2009)}]{dermer09_book}
{Dermer}, C.~D., \& {Menon}, G. 2009, {High Energy Radiation from Black Holes:
  Gamma Rays, Cosmic Rays, and Neutrinos}

\bibitem[{{Dermer} \& {Schlickeiser}(1993)}]{dermer93}
{Dermer}, C.~D., \& {Schlickeiser}, R. 1993, \apj, 416, 458

\bibitem[{{Dermer} \& {Schlickeiser}(2002)}]{dermer02}
---. 2002, \apj, 575, 667

\bibitem[{{Donnarumma} {et~al.}(2013){Donnarumma}, {Tramacere}, {Turriziani},
  {Costamante}, {Campana}, {De Rosa}, \& {Bozzo}}]{donnarumma13}
{Donnarumma}, I., {Tramacere}, A., {Turriziani}, S., {Costamante}, L.,
  {Campana}, R., {De Rosa}, A., \& {Bozzo}, E. 2013, ArXiv:1310.6965

\bibitem[{{Dotson} {et~al.}(2012){Dotson}, {Georganopoulos}, {Kazanas}, \&
  {Perlman}}]{dotson12}
{Dotson}, A., {Georganopoulos}, M., {Kazanas}, D., \& {Perlman}, E.~S. 2012,
  \apjl, 758, L15

\bibitem[{{Edelson} {et~al.}(2013){Edelson}, {Mushotzky}, {Vaughan}, {Scargle},
  {Gandhi}, {Malkan}, \& {Baumgartner}}]{edelson13}
{Edelson}, R., {Mushotzky}, R., {Vaughan}, S., {Scargle}, J., {Gandhi}, P.,
  {Malkan}, M., \& {Baumgartner}, W. 2013, \apj, 766, 16

\bibitem[{{Espaillat} {et~al.}(2008){Espaillat}, {Bregman}, {Hughes}, \&
  {Lloyd-Davies}}]{espaillat08}
{Espaillat}, C., {Bregman}, J., {Hughes}, P., \& {Lloyd-Davies}, E. 2008, \apj,
  679, 182

\bibitem[{{Finke}(2013)}]{finke13}
{Finke}, J.~D. 2013, \apj, 763, 134

\bibitem[{{Finke} {et~al.}(2008){Finke}, {Dermer}, \&
  {B{\"o}ttcher}}]{finke08_SSC}
{Finke}, J.~D., {Dermer}, C.~D., \& {B{\"o}ttcher}, M. 2008, \apj, 686, 181

\bibitem[{{Gaur} {et~al.}(2010){Gaur}, {Gupta}, {Lachowicz}, \&
  {Wiita}}]{gaur10}
{Gaur}, H., {Gupta}, A.~C., {Lachowicz}, P., \& {Wiita}, P.~J. 2010, \apj, 718,
  279

\bibitem[{{Ghisellini} {et~al.}(2011){Ghisellini}, {Tavecchio}, {Foschini}, \&
  {Ghirlanda}}]{ghisellini11_transition}
{Ghisellini}, G., {Tavecchio}, F., {Foschini}, L., \& {Ghirlanda}, G. 2011,
  \mnras, 414, 2674

\bibitem[{{Gonz{\'a}lez-Mart{\'{\i}}n} \& {Vaughan}(2012)}]{gonzalez12}
{Gonz{\'a}lez-Mart{\'{\i}}n}, O., \& {Vaughan}, S. 2012, \aap, 544, A80

\bibitem[{{Gupta} {et~al.}(2009){Gupta}, {Srivastava}, \& {Wiita}}]{gupta09}
{Gupta}, A.~C., {Srivastava}, A.~K., \& {Wiita}, P.~J. 2009, \apj, 690, 216

\bibitem[{{Hayashida} {et~al.}(2012){Hayashida}, {Madejski}, {Nalewajko},
  {Sikora}, {Wehrle}, {Ogle}, {Collmar}, {Larsson}, {Fukazawa}, {Itoh},
  {Chiang}, {Stawarz}, {Blandford}, {Richards}, {Max-Moerbeck}, {Readhead},
  {Buehler}, {Cavazzuti}, {Ciprini}, {Gehrels}, {Reimer}, {Szostek}, {Tanaka},
  {Tosti}, {Uchiyama}, {Kawabata}, {Kino}, {Sakimoto}, {Sasada}, {Sato},
  {Uemura}, {Yamanaka}, {Greiner}, {Kruehler}, {Rossi}, {Macquart}, {Bock},
  {Villata}, {Raiteri}, {Agudo}, {Aller}, {Aller}, {Arkharov}, {Bach},
  {Ben{\'{\i}}tez}, {Berdyugin}, {Blinov}, {Blumenthal}, {B{\"o}ttcher},
  {Buemi}, {Carosati}, {Chen}, {Di Paola}, {Dolci}, {Efimova}, {Forn{\'e}},
  {G{\'o}mez}, {Gurwell}, {Heidt}, {Hiriart}, {Jordan}, {Jorstad}, {Joshi},
  {Kimeridze}, {Konstantinova}, {Kopatskaya}, {Koptelova}, {Kurtanidze},
  {L{\"a}hteenm{\"a}ki}, {Lamerato}, {Larionov}, {Larionova}, {Larionova},
  {Leto}, {Lindfors}, {Marscher}, {McHardy}, {Molina}, {Morozova},
  {Nikolashvili}, {Nilsson}, {Reinthal}, {Roustazadeh}, {Sakamoto}, {Sigua},
  {Sillanp{\"a}{\"a}}, {Takalo}, {Tammi}, {Taylor}, {Tornikoski}, {Trigilio},
  {Troitsky}, \& {Umana}}]{hayashida12}
{Hayashida}, M., {et~al.} 2012, \apj, 754, 114

\bibitem[{{Joshi} \& {B{\"o}ttcher}(2007)}]{joshi07}
{Joshi}, M., \& {B{\"o}ttcher}, M. 2007, \apj, 662, 884

\bibitem[{{Kataoka} {et~al.}(1999){Kataoka}, {Mattox}, {Quinn}, {Kubo},
  {Makino}, {Takahashi}, {Inoue}, {Hartman}, {Madejski}, {Sreekumar}, \&
  {Wagner}}]{kataoka99}
{Kataoka}, J., {et~al.} 1999, \apj, 514, 138

\bibitem[{{Kataoka} {et~al.}(2001){Kataoka}, {Takahashi}, {Wagner}, {Iyomoto},
  {Edwards}, {Hayashida}, {Inoue}, {Madejski}, {Takahara}, {Tanihata}, \&
  {Kawai}}]{kataoka01}
---. 2001, \apj, 560, 659

\bibitem[{{Konigl}(1981)}]{konigl81}
{Konigl}, A. 1981, \apj, 243, 700

\bibitem[{{Kroon} \& {Becker}(2014)}]{kroon14}
{Kroon}, J.~J., \& {Becker}, P.~A. 2014, \apjl, 785, L34

\bibitem[{{Lachowicz} {et~al.}(2009){Lachowicz}, {Gupta}, {Gaur}, \&
  {Wiita}}]{lachowicz09}
{Lachowicz}, P., {Gupta}, A.~C., {Gaur}, H., \& {Wiita}, P.~J. 2009, \aap, 506,
  L17

\bibitem[{{Li} \& {Kusunose}(2000)}]{li00}
{Li}, H., \& {Kusunose}, M. 2000, \apj, 536, 729

\bibitem[{{Lott} {et~al.}(2012){Lott}, {Escande}, {Larsson}, \&
  {Ballet}}]{lott12}
{Lott}, B., {Escande}, L., {Larsson}, S., \& {Ballet}, J. 2012, \aap, 544, A6

\bibitem[{{Malmrose} {et~al.}(2011){Malmrose}, {Marscher}, {Jorstad},
  {Nikutta}, \& {Elitzur}}]{malmrose11}
{Malmrose}, M.~P., {Marscher}, A.~P., {Jorstad}, S.~G., {Nikutta}, R., \&
  {Elitzur}, M. 2011, \apj, 732, 116

\bibitem[{{Mastichiadis} {et~al.}(2013){Mastichiadis}, {Petropoulou}, \&
  {Dimitrakoudis}}]{mastichiadas13}
{Mastichiadis}, A., {Petropoulou}, M., \& {Dimitrakoudis}, S. 2013, \mnras,
  434, 2684

\bibitem[{{McHardy}(2008)}]{mchardy08}
{McHardy}, I. 2008, in Blazar Variability across the Electromagnetic Spectrum

\bibitem[{{Moderski} {et~al.}(2005){Moderski}, {Sikora}, {Coppi}, \&
  {Aharonian}}]{moderski05}
{Moderski}, R., {Sikora}, M., {Coppi}, P.~S., \& {Aharonian}, F. 2005, \mnras,
  363, 954

\bibitem[{{Mohan} {et~al.}(2011){Mohan}, {Mangalam}, {Chand}, \&
  {Gupta}}]{mohan11}
{Mohan}, P., {Mangalam}, A., {Chand}, H., \& {Gupta}, A.~C. 2011, Journal of
  Astrophysics and Astronomy, 32, 117

\bibitem[{{Nakagawa} \& {Mori}(2013)}]{nakagawa13}
{Nakagawa}, K., \& {Mori}, M. 2013, \apj, 773, 177

\bibitem[{{Nolan} {et~al.}(2012){Nolan}, {Abdo}, {Ackermann}, {Ajello},
  {Allafort}, {Antolini}, {Atwood}, {Axelsson}, {Baldini}, {Ballet}, \&
  et~al.}]{nolan12_2fgl}
{Nolan}, P.~L., {et~al.} 2012, \apjs, 199, 31

\bibitem[{{Rani} {et~al.}(2010){Rani}, {Gupta}, {Joshi}, {Ganesh}, \&
  {Wiita}}]{rani10}
{Rani}, B., {Gupta}, A.~C., {Joshi}, U.~C., {Ganesh}, S., \& {Wiita}, P.~J.
  2010, \apjl, 719, L153

\bibitem[{{Sbarrato} {et~al.}(2012){Sbarrato}, {Ghisellini}, {Maraschi}, \&
  {Colpi}}]{sbarrato12_transition}
{Sbarrato}, T., {Ghisellini}, G., {Maraschi}, L., \& {Colpi}, M. 2012, \mnras,
  421, 1764

\bibitem[{{Schlickeiser}(2009)}]{schlickeiser09}
{Schlickeiser}, R. 2009, \mnras, 398, 1483

\bibitem[{{Schlickeiser} {et~al.}(2010){Schlickeiser}, {B{\"o}ttcher}, \&
  {Menzler}}]{schlickeiser10}
{Schlickeiser}, R., {B{\"o}ttcher}, M., \& {Menzler}, U. 2010, \aap, 519, A9+

\bibitem[{{Sikora} {et~al.}(1994){Sikora}, {Begelman}, \& {Rees}}]{sikora94}
{Sikora}, M., {Begelman}, M.~C., \& {Rees}, M.~J. 1994, \apj, 421, 153

\bibitem[{{Sikora} {et~al.}(2009){Sikora}, {Stawarz}, {Moderski}, {Nalewajko},
  \& {Madejski}}]{sikora09}
{Sikora}, M., {Stawarz}, {\L}., {Moderski}, R., {Nalewajko}, K., \& {Madejski},
  G.~M. 2009, \apj, 704, 38

\bibitem[{{Sobolewska} {et~al.}(2014){Sobolewska}, {Siemiginowska}, {Kelly}, \&
  {Nalewajko}}]{sobolewska14}
{Sobolewska}, M.~A., {Siemiginowska}, A., {Kelly}, B.~C., \& {Nalewajko}, K.
  2014, \apj, in press, arXiv:1403.5276

\bibitem[{{Tanaka} {et~al.}(2013){Tanaka}, {Cheung}, {Inoue}, {Stawarz},
  {Ajello}, {Dermer}, {Wood}, {Chekhtman}, {Fukazawa}, {Mizuno}, {Ohno},
  {Paneque}, \& {Thompson}}]{tanaka13}
{Tanaka}, Y.~T., {et~al.} 2013, \apjl, 777, L18

\bibitem[{{Vaughan}(2005)}]{vaughan05}
{Vaughan}, S. 2005, \aap, 431, 391

\bibitem[{{Vaughan} \& {Uttley}(2006)}]{vaughan06}
{Vaughan}, S., \& {Uttley}, P. 2006, Advances in Space Research, 38, 1405

\bibitem[{{Wehrle} {et~al.}(2013){Wehrle}, {Wiita}, {Unwin}, {Di Lorenzo},
  {Revalski}, {Silano}, \& {Sprague}}]{wehrle13}
{Wehrle}, A.~E., {Wiita}, P.~J., {Unwin}, S.~C., {Di Lorenzo}, P., {Revalski},
  M., {Silano}, D., \& {Sprague}, D. 2013, \apj, 773, 89

\bibitem[{{Zacharias} \& {Schlickeiser}(2010)}]{zacharias10}
{Zacharias}, M., \& {Schlickeiser}, R. 2010, \aap, 524, A31+

\bibitem[{{Zacharias} \& {Schlickeiser}(2012{\natexlab{a}})}]{zacharias12}
---. 2012{\natexlab{a}}, \mnras, 420, 84

\bibitem[{{Zacharias} \& {Schlickeiser}(2012{\natexlab{b}})}]{zacharias12_EC}
---. 2012{\natexlab{b}}, \apj, 761, 110

\bibitem[{{Zacharias} \& {Schlickeiser}(2013)}]{zacharias13}
---. 2013, \apj, 777, 109

\bibitem[{{Zhang}(2002)}]{zhang02_mrk421}
{Zhang}, Y.~H. 2002, \mnras, 337, 609

\bibitem[{{Zhang} {et~al.}(1999){Zhang}, {Celotti}, {Treves}, {Chiappetti},
  {Ghisellini}, {Maraschi}, {Pian}, {Tagliaferri}, {Tavecchio}, \&
  {Urry}}]{zhang99}
{Zhang}, Y.~H., {et~al.} 1999, \apj, 527, 719

\bibitem[{{Zhang} {et~al.}(2002){Zhang}, {Treves}, {Celotti}, {Chiappetti},
  {Fossati}, {Ghisellini}, {Maraschi}, {Pian}, {Tagliaferri}, \&
  {Tavecchio}}]{zhang02_2155}
---. 2002, \apj, 572, 762

\end{thebibliography}

\end{document}